\documentclass[aps,pre,twocolumn,groupedaddress]{revtex4-1}
\usepackage{graphicx}
\usepackage{subfigure}
\usepackage{amsmath}

\usepackage[usenames, dvipsnames]{color}

\begin{document}

%\preprint{}

\title{Effects of Initial Condition Spectral Content on Shock Driven Turbulent Mixing}


\author{Nicholas J. Nelson}
\email[]{njnelson@lanl.gov}
%\thanks{}
\affiliation{Los Alamos National Laboratory, Los Alamos, NM 87545}

\author{Fernando F. Grinstein}
%\email[]{}
%\thanks{}
\affiliation{Los Alamos National Laboratory, Los Alamos, NM 87545}


\date{\today}

\begin{abstract}
The mixing of materials due to the Richtmyer-Meshkov instability and the ensuing turbulent behavior is of intense interest in a variety of physical systems including inertial confinement fusion, combustion, and the final stages of stellar evolution. Extensive numerical and laboratory studies of shock-driven mixing have demonstrated the rich behavior associated with the onset of turbulence due to the shocks. Here we report on progress in understanding shock-driven mixing at interfaces between fluids of differing densities through 3D numerical simulations using the RAGE code in the implicit large eddy simulation context. We consider a shock tube configuration with a band of high density gas (SF$_6$) embedded in low density gas (air). Shocks with a Mach number of 1.26 are passed through SF$_6$ bands, resulting in transition to turbulence driven by  the Richtmyer-Meshkov instability. The system is followed as a rarefaction wave and a reflected secondary shock from the back wall pass through the SF$_6$ band. We apply a variety of initial perturbations to the interfaces between the two fluids in which the physical standard deviation, wave number range, and the spectral slope of the perturbations are held constant, but the number of modes initially present is varied. By thus decreasing the density of initial spectral modes of the interface, we find that we can achieve as much as 25\% less total mixing at late times. This has potential direct implications for the treatment of initial conditions applied to material interfaces in both 3D and reduced dimensionality simulation models.
\end{abstract}

\pacs{}
%\keywords{}

\maketitle

\section{Introduction \label{sec:Intro}}

Mixing of materials due to turbulent behavior has been a long-standing problem of interest in many areas of physics. In particular, the growth of interface perturbations due to impulsive driving of an interface between materials has been of great interest in diverse subjects including inertial confinement fusion \cite{Haines2014}, relativistic astrophysical jets \cite{Matsumoto2013}, and supernovae remnants \cite{Inoue2013}. All of these subjects include mixing across interfaces due to shock-driven turbulence. The Richtmyer-Meshkov instability (RMI) \cite{Brouillette2002} drives the growth of initial perturbations until amplitudes become sufficient to drive nonlinear interactions between modes. This coupling of modes drives a transition from laminar to turbulent behavior and thus allows turbulent mixing of the two fluids. RMI adds the challenges associated with shock physics and transitional turbulence to the already difficult problem of hydrodynamical mixing.

Small-scale resolution requirements for simulations typically focus on those of continuum
fluid mechanics described by the Navier-Stokes equations; different requirements are
involved depending on the regime considered and on the relative importance of coupled
physics such as multi-species diffusion and combustion as determined by Knudsen, Reynolds,
Schmidt, Damkohler, and other characteristic non-dimensional numbers; on the other hand,
the longest wavelengths that can be resolved are constrained by the size of the
computational domain. Ideally, we would like to resolve all relevant space/time scales and
material interfaces in our simulations, the so-called direct numerical simulation (DNS)
strategy. DNS is prohibitively expensive in the foreseeable future for most practical flows and regimes of interest at moderate-to-high
Reynolds number (Re). On the other end of the simulation spectrum are the Reynolds-Averaged Navier-Stokes (RANS) 
approaches which are the preferred industrial standard. In coarse grained simulation (CGS) strategies, large energy containing structures are resolved, smaller structures are
filtered out, and unresolved sub-grid scale (SGS) effects are modeled; this includes
classical large eddy simulation (LES) strategies with explicit use of SGS models, \cite{Sagaut2006} and implicit LES
(ILES) \cite{Grinstein2010}, relying on SGS modeling and filtering provided by physics capturing
numerical algorithms. The CGS strategy of separating resolved and SGS physics effectively
becomes the intermediate approach between DNS and RANS.

Turbulent material mixing can be usefully characterized by the fluid physics involved: large-scale entrainment,
stirring due to velocity gradient fluctuations, and, molecular diffusion. At moderately
high Re, when convective time-scales are much smaller than those associated with
molecular diffusion, we are primarily concerned with the numerical simulation of the first
two convectively-driven processes. These processes can be captured with sufficiently
resolved ILES \cite{Grinstein2010}, which serves as our primary simulation strategy. 

While the RMI has been studied in a wide variety of geometries \cite{Bates2007, Thornber2010, Grinstein2011, Haines2013}, here we will focus on a shock-tube configuration with a band of high density gas (sodium hexafluoride SF$_6$) embedded in a low density gas (air). This will allow us to study mixing from the initial shock, the rarefaction wave, 
and reshock after primary shock reflection off the back wall of the system.

In instability-driven turbulence, special attention has been paid to the initial conditions (IC) which seed the ballistic and nonlinear phases of instability growth and the transition to turbulent behavior. Extensive numerical and laboratory studies have shown that the energetic and mixing properties of RMI-induced flows carry significant imprint from their IC even at late times.   Beyond the work surveyed in \cite{Brouillette2002}, investigation of IC effects on RMI have been subject of many experimental, \cite{VS, Jacobs2013, Poggi, Balasubramanian2012, Leinov} computational, \cite{Cohen2002, Hill2006, Leinov, Schilling, Youngs, Hahn, Thornber2010, Ukai, Grinstein2011} as well as theoretical studies. \cite{Thornber2010, Mikaelian}.  These investigations follow on the current recognition that depending on the IC, different far field or late time self-similar regimes are possible \cite{George2004}. In particular, computational  \cite{Grinstein2011} and laboratory \cite{Balasubramanian2012} experiments show that the addition of high-wave number modes to the initial material interface perturbations can greatly increase late time mixing after reshock.

In this paper, we extend this investigation of the dependence of late-time behavior on IC by modeling the impact of a Mach number 1.26 shock on a 15 cm wide band of SF$_6$. For each simulation, we apply initial perturbations to both the front and back surfaces of the SF$_6$ band. These perturbations have the same physical standard deviation, range of wave numbers, and spectral slope in all simulations presented in this paper. We examine the effect of varying the initial spectral density by modifying spectral content over a given band of wave numbers using fractions of the total number of modes possible. We find that in cases with fewer initial Fourier modes we can show as much as 25\% less mixing at late times compared to simulations with a full spectrum of initial perturbations. We attribute this variation to the hinderance of the development of a turbulent cascade.

\subsection{Guidance from Laboratory Experiments}

Laboratory studies of RMI-driven mixing provide crucial guidance for the simulations which we present in this work. A rich history of experimental work has shown that RMI-induced flows and mixing are strongly dependent on the shock velocity, the geometry of the system, and the initial interface perturbations \cite{Prestridge2013}. Extensive effort has yielded a wide variety of measurement techniques which allow the extraction of rich datasets from experiments, including 2D concentration maps which enable measurements of the mixing rate \cite{Weber2012}, and simultaneous concentration and velocity field diagnostics for detailed analysis of turbulent mixing \cite{Tomkins2013}. 

With advanced diagnostic tools, laboratory experiments have been designed to study the impact of shock velocity and IC on the mixing properties of RMI-driven turbulence. IC have been shown to play a significant role in the development of the flow and mixing properties of experiments. In shock-tube experiments, inconsistencies between realizations has been attributed to variations in IC \cite{Balasubramanian2012, Jacobs2013}. These studies provide important general trends against which we can compare our results.

\subsection{Numerical Investigations}

Simulations of RMI-induced mixing are fundamentally limited by the nature of shock discontinuities. Without the possibility of direct numerical simulations (DNS), 3D models are limited to a large-eddy simulation (LES) approach in which large scales are computationally resolved and the effects of unresolved scales is modeled either explicitly or by the choice of highly-stable numerical operators. In somewhat similar fashion, laboratory experiments are limited by the finite scales of the instrumental resolution of measuring/visualizing devices characterizing the resolution of the observational techniques \cite{Grinstein2009}. 

In the LES framework, unresolved scales are generally assigned a diffusive character and represented by a suitable diffusive operator. One option is to add explicit diffusive terms which represent unresolved diffusive processes. A wide range of such explicit operators have been studied in a variety of engineering, geophysical, and astrophysical settings \cite{Meneveau2000, Buffett2003, Kirkpatrick2006, Rempel2009, Nelson2013a}. 

Classical LES is particularly inadequate for flows driven by RMI because of the dissipative numerics needed for shock capturing \cite{Sagaut2006}. Hybrid methods switching between shock-capturing schemes and conventional LES depending on the local flow conditions have been proposed \cite{Hill2006}. Such high-order shock-capturing (e.g., fifth-order WENO) methods are typically chosen to ensure a smooth transition and matching of the inherently different simulation models. However, all shock capturing methods degrade to first-order in the vicinity of shocks because of the monotonicity requirements Ð and in particular, at the very important initial stage at which a shock first interacts with a compositional interface and generate the fluctuating velocity field. Thus, severe resolution demands to address the competition between the implicit subgrid models provided by numerical operators and explicit subgrid models can be expected in the hybrid context.  Alternatively, many codes now employ so-called physics-capturing numerical schemes which by design rely on minimal numerical diffusion for computational stability. This ILES framework permits simulations to run with minimal levels of diffusion which is crucial for correctly modeling shock physics \cite{Grinstein2010}.  By combining shock and turbulence emulation capabilities based on a single physics-capturing numerical model, ILES  provides a natural effective simulation strategy for RMI \cite{Dri2005}. Moreover, ILES can accurately capture the stirring-driven scalar mixing which is the dominant aspect for high Re flows \cite{Wachtor2013}.

%
%  Here's the new Wachtor2013 reference:  A.J. Wachtor, F.F. Grinstein, C.R. DeVore, J.R. Ristorcelli, and L.G. Margolin, Physics of Fluids, 25, 025101, 2013.
%

 \begin{figure*}
 \includegraphics[width=0.6\linewidth]{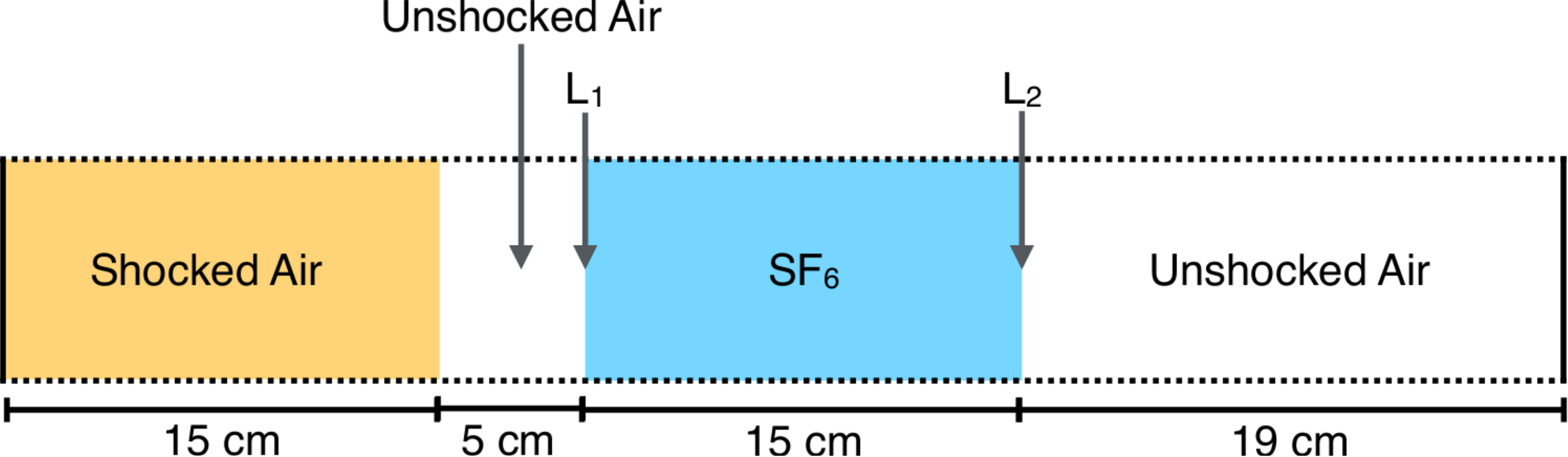}
 \caption{(Color online) Schematic diagram of the IC for all simulations (see Table~\ref{table:Cases}). The shock wave initially propagates to the right. Solid lines represent solid boundaries at both ends of the shock-tube, while dotted lines show periodic boundaries in the transverse directions. Initial perturbations are applied to the front and back interfaces of the SF$_6$ band, labeled L$_1$ and L$_2$ respectively.  \label{fig:Setup}}
 \end{figure*}

Extensive numerical studies have investigated many properties of the RMI in a variety of geometries, including the single planar interface \cite{Cohen2002, Hill2006, Leinov, Schilling, Youngs, Hahn, Thornber2010, Ukai, Grinstein2011,  Ristorcelli2013}, the two-interface inverse chevron \cite{Hahn, Haines2013}, the shocked heavy cylinder \cite{Weirs}, and the shocked gas curtain \cite{Gowardhan2011a}.  Special attention has been paid to the effects of IC specifics \cite{Grinstein2011, Ristorcelli2013}. ILES codes have been also used in a wide variety of hydrodynamic and magnetohydrodynamic applications including airfoil design, planetary magnetospheres, inertial confinement fusion, and stellar dynamo models, e.g.,\cite{Barnes2013, Zhu2014, Haines2014, Racine2011}.

%\subsection{}
%\subsubsection{}

% If in two-column mode, this environment will change to single-column
% format so that long equations can be displayed. Use
%\begin{widetext}
%\end{widetext}

% figures should be put into the text as floats.
% Use the graphics or graphicx packages (distributed with LaTeX2e)
% and the \includegraphics macro defined in those packages.
% See the LaTeX Graphics Companion by Michel Goosens, Sebastian Rahtz,
% entire page. There is no need to do explicit centering.

\section{Computational Tools and Simulation Parameters \label{sec:Setup}}

Here we use the ILES capability of the code RAGE \cite{Gittings2008} to study the RMI and associated induced turbulent mixing. RAGE has been used extensively for similar simulations, reporting good agreement with available laboratory observations \cite{Grinstein2011, Gowardhan2011a, Haines2013, Haines2013a, Haines2014}.  RAGE solves the Euler equations of multi-fluid compressible hydrodynamics using a secondÐorder Godunov finite-volume scheme, and offering a variety of numerical options. In the present simulations, adaptive mesh refinement and material interface RAGE options were not used; a Van Leer limiter and uniform Cartesian gridding were chosen.

The evolution equations encode conservation of mass, material concentrations, momentum, and energy given by
\begin{equation}
\frac{ \partial \rho }{ \partial t } + \nabla \cdot \left( \rho \vec{u} \right) = 0
\label{eq:mass}
\end{equation}
\begin{equation}
\frac{ \partial \rho Y_n }{ \partial t } + \nabla \cdot \left( \rho Y_n \vec{u} \right) = 0
\label{eq:concentration}
\end{equation}
\begin{equation}
\frac{ \partial \rho \vec{u} }{ \partial t } + \left( \vec{u} \cdot \nabla \right) \left( \rho \vec{u} \right) + \nabla P = 0
\label{eq:momentum}
\end{equation}
\begin{equation}
\frac{ \partial \rho E }{ \partial t } + \nabla \cdot \left[ \left(\rho E + P \right) \vec{u} \right] = 0 ,
\label{eq:energy}
\end{equation}
where $\rho$ is the mass density, $\vec{u}$ is the velocity, $Y_n$ is the mass fraction of species $n$, $P$ is the fluid pressure, and $E$ is the total specific energy.  These equations are supplemented with SESAME tabular equations of state \cite{Lyons1992}.

%\bibitem{sesame}
%S.~P.~Lyon \& J.~D.~Johnson.
%\newblock SESAME: the Los Alamos National Laboratory Equation of State Database.
%\newblock {\em Los Alamos National Laboratory} LA-UR-92-3407, 1992.

We define the transverse average of a quantity $f (x,y,z) $ as
\begin{equation}
 \bar{f}(x) = \frac{ \int \int f \, dy \, dz}{ L_y L_z } 
 \end{equation}
 and the mass-weighted transverse mean as
\begin{equation}
 \tilde{f}(x) = \frac{ \int \int \rho f \, dy \, dz}{ \bar{\rho} L_y L_z } .
 \end{equation}
We have neglected body forces, molecular shear, and heat conduction.  The ratio of specific heats for mixtures is computed using a volume-fraction weighted average of the ratio of specific heats for each gas. As used in the present work, RAGE models miscible material interfaces (Schmidt number $Sc \sim 1$) and high Re convection-driven flow with effective numerical viscosity determined by the small-scale cutoff.  Timesteps in RAGE runs are set as minimum of a Courant condition on the hydrodynamics and stability criteria determined by other physics in the code.

\begin{figure*}
 \subfigure[Initial Surface Perturbations for Case S-1]{\includegraphics[width=0.43\linewidth]{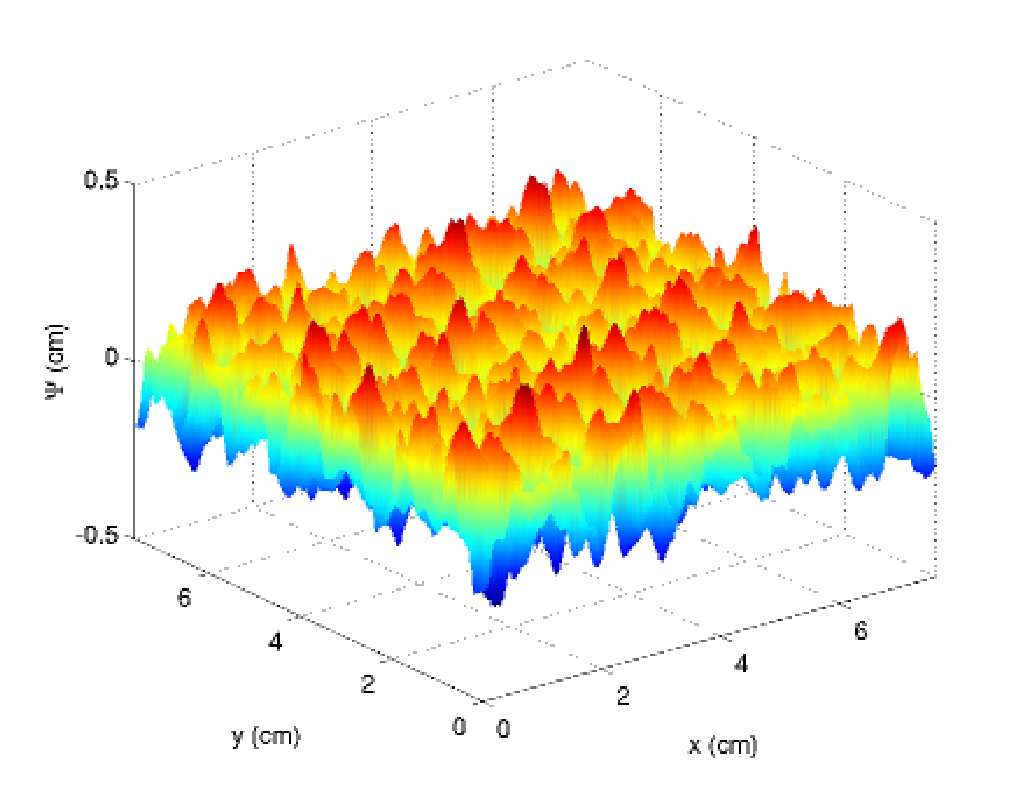}}
  \subfigure[Spectral Content of (a)]{\includegraphics[width=0.43\linewidth]{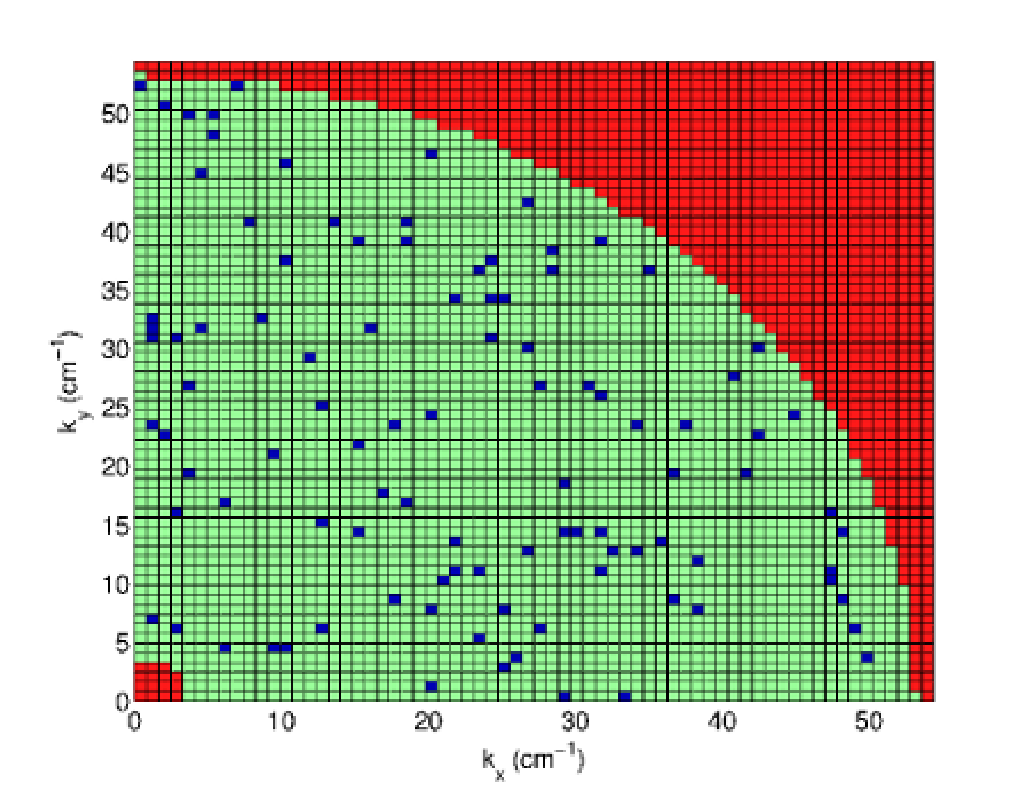}}
    \subfigure[Initial Surface Perturbations for Case S-8]{\includegraphics[width=0.43\linewidth]{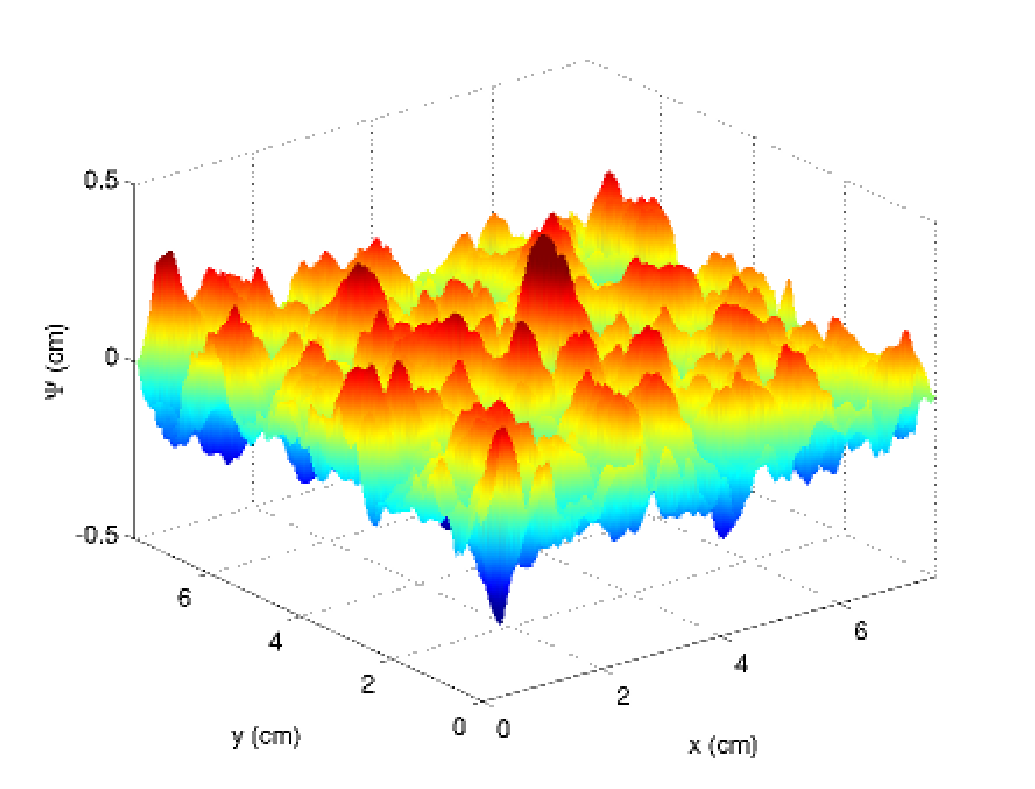}}
\subfigure[Spectral Content of (c)]{\includegraphics[width=0.43\linewidth]{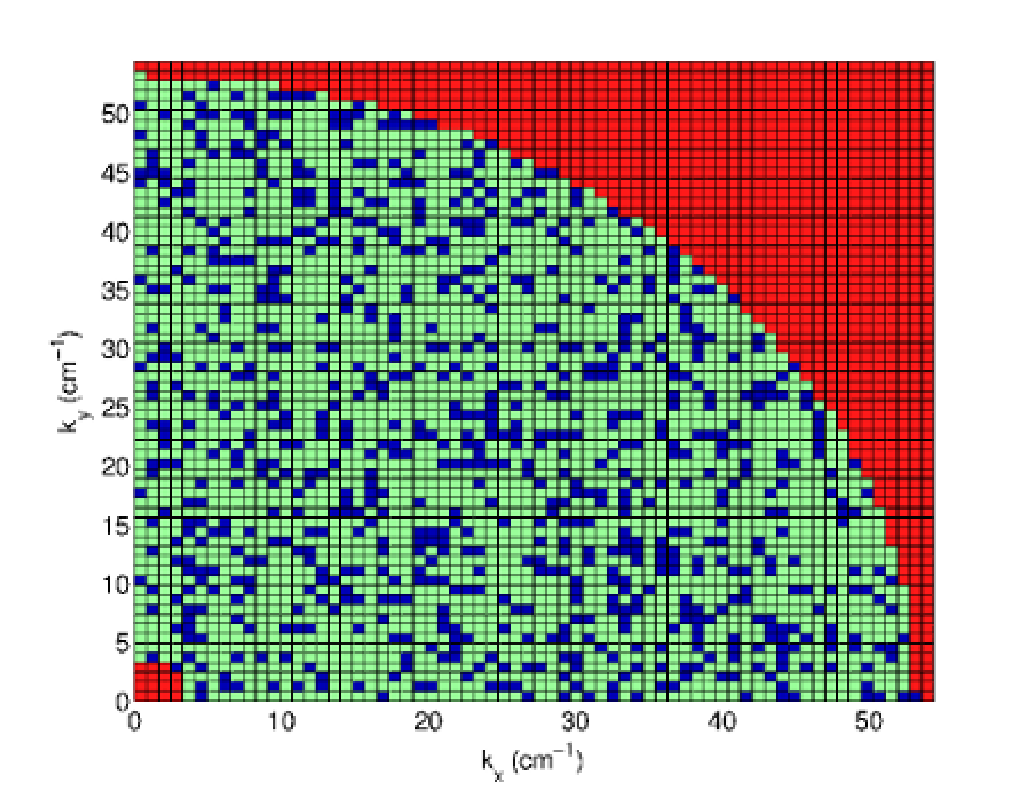}}
 \subfigure[Initial Surface Perturbations for Case S-32]{\includegraphics[width=0.43\linewidth]{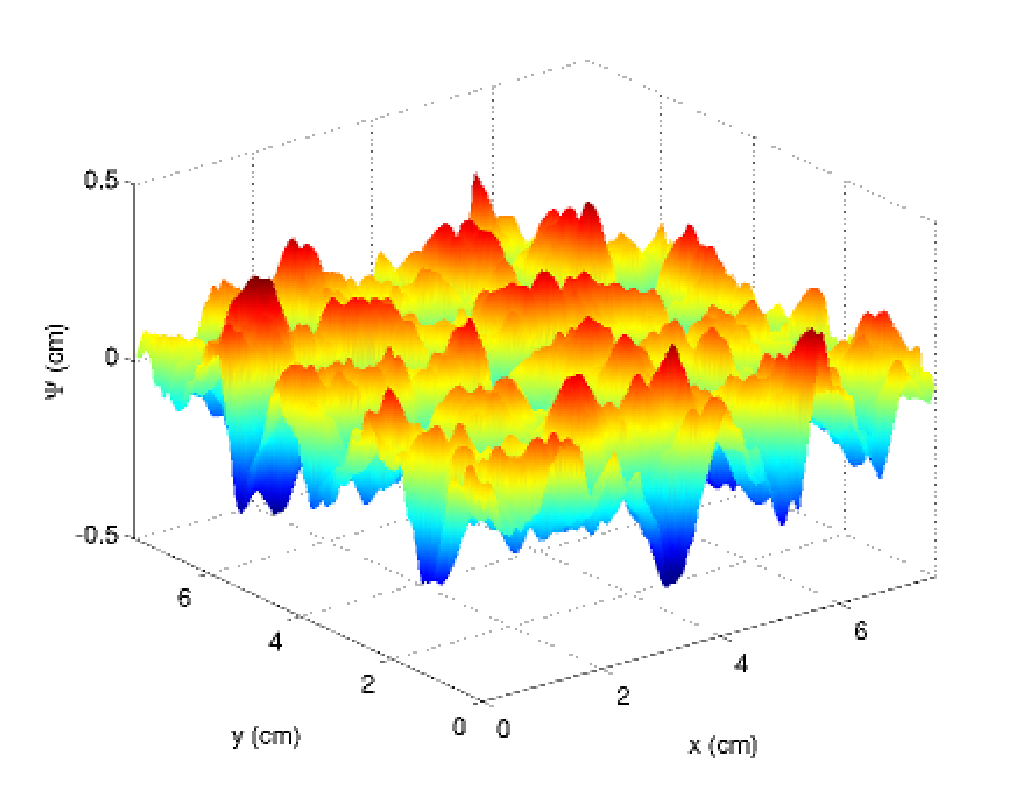}}
   \subfigure[Spectral Content of (e)]{\includegraphics[width=0.43\linewidth]{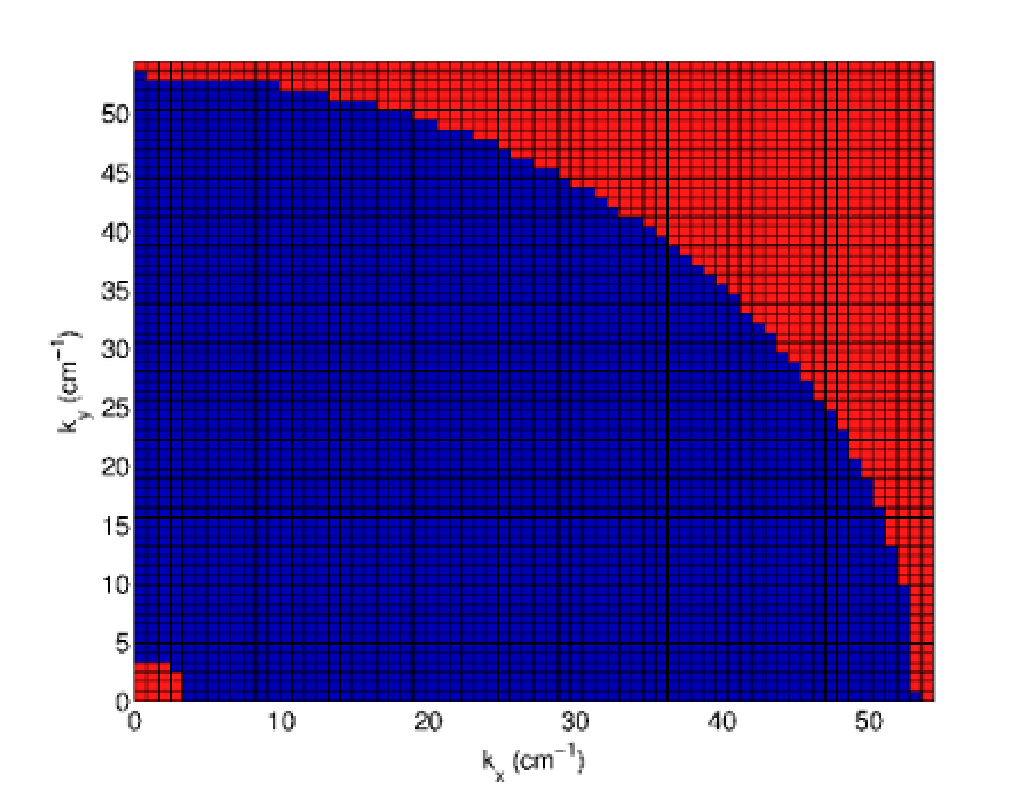}}
 \caption{(Color online) Initial surface perturbations with (a) 100 modes, (b) 800 modes, and (c) 3263 modes, along with the spectral content of those surfaces shown in (b), (d), and (f) for each case respectively. All surfaces have a standard deviation of 0.1 cm and random phasing of the modes. For (b), (d), and (f), red (medium gray) indicates modes with wave numbers that do not satisfy Equation~\ref{eq:modeNums}, green (light gray) indicates eligible modes that are not used in this realization of the interface surface, and blue shows (dark gray) modes which have been used to construct the associated surface. Note that in (f) all modes compliant with Equation~\ref{eq:modeNums} are used. \label{fig:InitPert}}
 \end{figure*}
 
Here we report on a series of nine simulations which all follow the basic physical setup shown in Figure~\ref{fig:Setup}. The simulation domain is 54 cm long by 7.62 cm square in the transverse direction. At the start of the simulation shocked air fills the leftmost 15 cm of the domain, with a 5 cm gap between the shock front and the first interface between the unshocked air and the SF$_6$, which we label $L_1$. The initially stationary SF$_6$ band has a width of 15 cm. The right side of the band is labeled the second interface between unshocked air and SF$_6$ or $L_2$. The reminder of the domain is filled with unshocked air. The Atwood number for both interfaces is 0.69. 

A shocked air region is created upstream in terms of a higher-density higher-pressure region chosen to satisfy the Rankine-Hugoniot relations for a Mach 1.26 shock. The primary shock propagates in the $x$-direction through the contact discontinuities and reflects at the end of the simulation box on the right. Boundary conditions involve reflecting walls on the right in the $x$-direction and periodicity in the transverse directions.

For this system, a simple estimate of the Reynolds number for the physical system can be found using the shock velocity, the transverse width of the domain, and the viscosity of air. This gives a Re of approximately $6 \times 10^6$. The enormous range of scales which would be needed make a true DNS calculation impractical for the foreseeable future. Thus we choose a CGS strategy -- RAGE based ILES. 

\subsection{Initial Interface Perturbations}

IC are applied to both interfaces as perturbations in the $x$ position of the interface at each point. Thus the perturbed interface location is given by $x + \psi (y,z)$, where $\psi (y,z)$ is the perturbation of the interface. We define $\psi$ as 
 %
%\begin{equation}
\begin{eqnarray}
\psi \left( y, z \right) = \sum_{i=1}^{N_{IC}} & A_i \left[ C_i \left( \cos \left( k_i y + l_i z \right) \right. \right. \nonumber \\ 
& + D_i \left. \sin \left( k_i y + l_i z \right) \right] \nonumber \\ 
\end{eqnarray}
%\begin{eqnarray}
%\psi \left( y, z \right) = & \sum_{i=1}^{N_{IC}} A_i \left[ C_i \left( \cos \left( k_i y \right) \cos \left( l_i z \right) - \sin \left( k_i y \right) \sin \left( l_i z \right) \right) \right. \nonumber \\ 
%& -  \left. D_m \left( \cos \left( k_i y \right) \sin \left( l_i z \right) + \sin \left( k_i y \right) \cos \left( l_i z \right) \right)  \right]
%\end{eqnarray}
% \end{equation}
 %
 where $N_{IC}$ is the number of modes to be applied as initial perturbations, $A_i$ is the amplitude of mode with wave number $k_i$ in the $y$-direction and $l_i$ in the $z$-direction. Coefficients $C_i$ and $D_i$ are randomly selected from a normal distribution with 0 mean and a variance of 1. This is equivalent to stating that the modes have purely random phases. Amplitudes $A_i$ are chosen such that $ A_i \propto \left( k_i^2 + l_i^2 \right)^{-1}$ so that the amplitude of a given mode scales as the magnitude of its wavevector to the $-2$ power. The values for all $A_i$ are uniformly scaled so that the standard deviation of the interface perturbations is 0.1 cm for all simulations. Different mode sets with identical standard deviation, spectral slope, and numbers of initial modes are used for the the front and back interfaces of the band to avoid any type of synchronization between the two interfaces.
 
 \begin{table}%[H] add [H] placement to break table across pages
\caption{Simulation parameters for primary models, including the numbers of grid points in the $x$ (streamwise), $y$, and $z$ directions $N_x \times N_y \times N_z $, the grid spacing $\Delta x$ in $\mu$m, and the number of modes used to construct the initial interface perturbations on either side of the SF$_6$ band, and the number of randomly generated realizations simulated $N_R$. \label{table:Cases}}
\begin{ruledtabular}
\begin{tabular}{lcccc}
Case & $N_x \times N_y \times N_z $ & $\Delta x$ ($\mu$m) &  $N_\mathrm{IC}$ & $N_R$  \\
\hline
V-1 & $4552 \times 600 \times 600 $ & 127 & 100 & 1 \\
H-32 & $2990 \times 422 \times 422 $ & 181 & 3263 & 1 \\
H-1 & $2990 \times 422 \times 422 $ & 181 & 100 & 1 \\
S-32 & $1800 \times 254 \times 254 $ & 300 & 3263 & 3 \\
S-16 & $1800 \times 254 \times 254 $ & 300 & 1600 & 1 \\
S-8 & $1800 \times 254 \times 254 $ & 300 & 800 & 1 \\
S-4 & $1800 \times 254 \times 254 $ & 300 & 400 & 3 \\
S-2 & $1800 \times 254 \times 254 $ & 300 & 200 & 1 \\
S-1 & $1800 \times 254 \times 254 $ & 300 & 100 & 5 \\
\end{tabular}
\end{ruledtabular}
\end{table}
 
Our broadband initial perturbations all utilize the same range of wave numbers. We select modes such that 
\begin{equation}
 3.298 \; \mathrm{cm}^{-1} \leq \sqrt{k_i^2 + l_i^2} \leq 52.77 \; \mathrm{cm}^{-1} 
 \end{equation}
or equivalently 
\begin{equation}
4 \leq \sqrt{m_i^2 + n_i^2} \leq 64
\label{eq:modeNums}
 \end{equation}
where $m_i$ and $n_i$ are the mode numbers in the $y$- and $z$-directions, respectively. This means that there are 3263 possible distinct wave number pairs that can generate orthogonal modes, which we can then apply as interface perturbations. For simulations with fewer than the maximum number of modes, we randomly select a number of modes $N_{IC}$ from the 3263 total possible. The selection process is done by choosing $N_{IC}$ random integers from a uniform distribution between 0 and 64 for $m_i$ and $n_i$ independently such that the constraint in Equation~\ref{eq:modeNums} is satisfied. We also do not allow duplicate modes, so accidentally occurring repeats of $\left( m_i, n_i \right)$ pairs are removed and reselected. In this way the modes correctly sample the mode number space with the same fraction of possible modes selected on average from any two bands in mode magnitude. For example, the interface perturbations used with $N_{IC} = 100$ had 10 modes with $ 10 \leq \sqrt{m_i^2 + n_i^2} \leq 20$ out of a possible 249 modes, or 3.7\% of all possible modes, and 15 out of 401 modes with mode magnitudes between 20 and 30, or 4.0\%. As expected, cases with $N_{IC} = 200$ utilize 6.8\% and 7.0\% of the possible modes in these two bands, respectively. The percentages of utilized modes are 12.9\% and 13.2\% for cases with $N_{IC} = 400$, 25.3\% and 24.2\% for cases with $N_{IC} = 800$, and 49.8\% and 49.4\% for cases with $N_{IC} = 1600$. Thus the fraction of possible modes utilized in any band in mode magnitude will be roughly equal to the overall fraction of possible modes used. We define this fraction of modes used over any range of values in mode magnitude as the spectral density of the surface. 

  \begin{figure}
 \includegraphics[width=\linewidth]{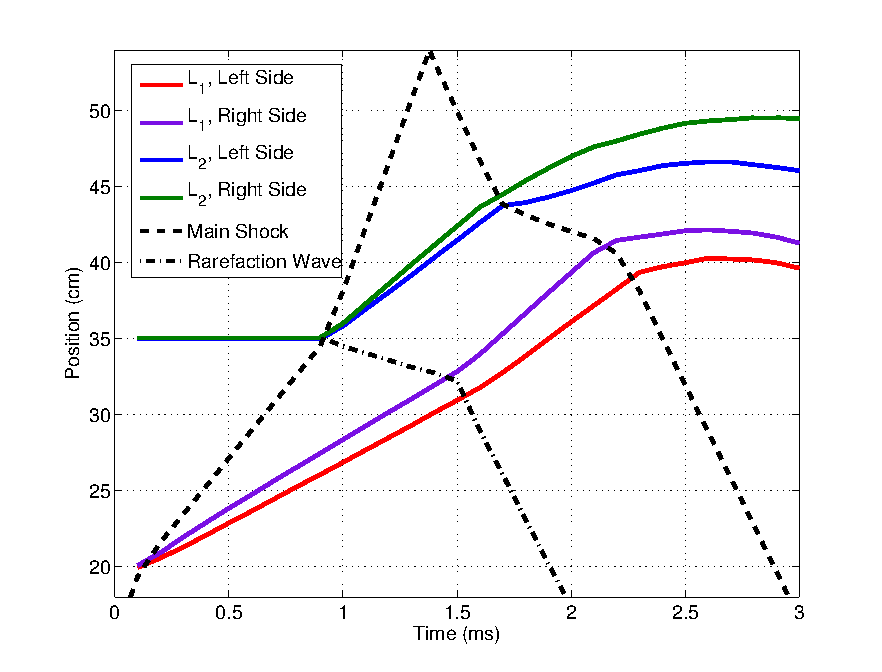}
 \caption{(Color online) A time-space diagram showing the position of the front (bottom solid line) and back (second solid line from bottom) of both mixing layers $L_1$ and $L_2$ (front is second solid line from top, back is top solid line) as well as the primary shock and the rarefaction wave for case S-32.  \label{fig:spacetime}}
 \end{figure}

   \begin{figure*}
 \includegraphics[width=\linewidth]{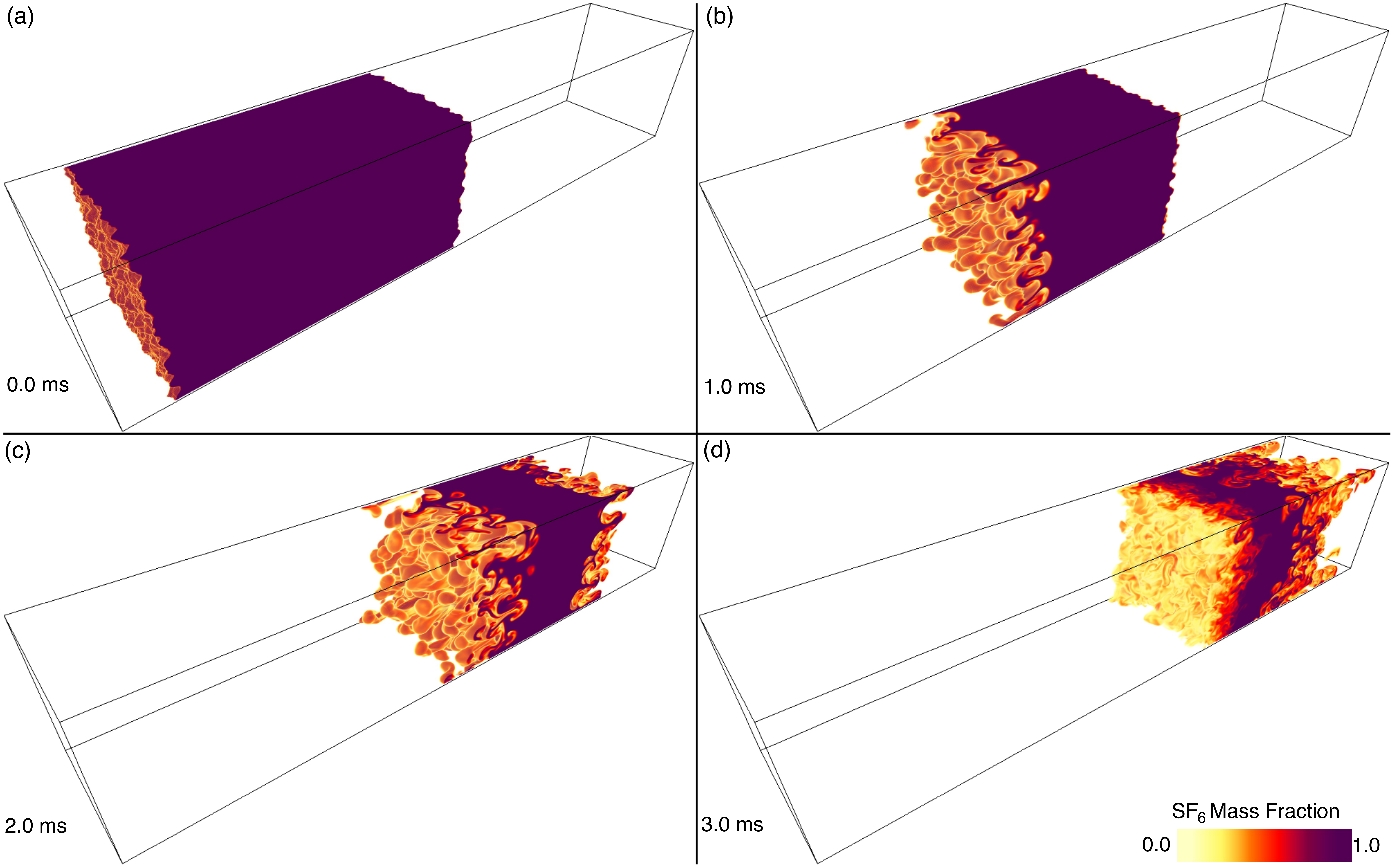}
 \caption{(Color online) Snapshot 3D volume renderings of SF$_6$ mass fraction from $x=18$ cm to $x = 54$ cm for case S-32 at (a) the start of the simulation, (b) $t = 1.0$ ms when the shock front is in the SF$_6$ band, (c) $t = 2.0$ ms when the shock has exited the band, and (d) $t = 3.0$ ms after reshock. $Y_\mathrm{SF6}$ is shown with low concentrations in yellow (light gray), high concentrations in purple (dark gray), and pure air transparent. Visualizations made with VAPOR \cite{Clyne2007}. \label{fig:Vapor_4times}}
 \end{figure*}

Figure~\ref{fig:InitPert} shows the initial interface perturbations applied to the front interface $L_1$ for simulations with 100, 800, and 3263 modes, as well as a schematic representation of the modes present in each interface. All interfaces have the same RMS amplitude, but visual differences can be noticed due to the differing spectral content. For example, comparing Figures~\ref{fig:InitPert}(a) and (e) reveals that the case with 32 times the number of modes shows considerably more spatial inhomogeneity. Quantitatively, the surface with $N_{IC} = 100$ has a kurtosis of 2.24 while the surface with $N_{IC} = 3263$ has a kurtosis of 3.02, or slightly more than Guassian. As the number of modes increases, the surface perturbations become more Gaussian and thus approach a kurtosis of three. 

Table~\ref{table:Cases} list the simulations discussed in this paper and their attributes. For all but cases V-1, H-32 and H-1, we use a grid spacing of 0.03 cm, giving a computational domain that is 1800 cells by 254 cells square for a total of 116 million computational elements in each simulation. Cases H-32 and H-1 use a grid spacing of 0.018 cm with the same physical domain for a total of 929 million computational elements. Case V-1 uses a grid spacing of 0.0127 cm for a total of 1.64 billion computational cells.

As both the modes selected and the phases of each individual mode are randomly assigned there is the potential for different realizations of our IC to produce significantly different results. We might expect this realization noise to be particularly pronounced for low numbers of initial modes where there may be only a few modes selected at the longest wavelengths. To address this issue, we have run multiple realizations of cases S-32, S-4, and S-1 in order to give some idea of the magnitude of the realization noise at these three values of $N_{IC}$ Ð discussed further below.

\subsection{Simulation Results}

All simulations were begun from IC like those illustrated in Figure~\ref{fig:Setup} and evolved for at least 3.0 ms. Figure~\ref{fig:spacetime} shows the evolution of case S-32 as a sample. The primary shock hits the first interface about 0.1 ms after the start of the simulation and begins compressing the band. The mixing layer at this interface begins to grow at that point. The primary shock hits the back interface at about $t = 0.92$ ms, launching a rarefaction wave and beginning the growth of the second mixing layer. The primary shock reflects off the back wall at about $t = 1.38$ ms. The rarefaction wave hits the first interface at about $t = 1.50$ ms and continues to propagate through the shocked air. The reflected primary shock hits the second interface at about $1.55$ ms and proceeds to reshock the band until it hits the first interface again at about $t = 2.15$ ms.

Following previous studies of shock-driven mixing \cite{Gowardhan2011, Gowardhan2011a, Grinstein2011, Haines2013}, we can describe the mixing layers at each interface in terms of the SF$_6$ mass fraction, $Y_{\mathrm{SF6}} $, using the mixing parameter $\phi$ defined as 
\begin{equation}
\phi = 4 Y_{\mathrm{SF6}} \left(1 - Y_{\mathrm{SF6}}  \right),
\label{eq:phi}
 \end{equation}
where $\phi$ ranges from zero for pure air or SF$_6$ to one for a 50/50 mix by mass fraction of both species.  In discussing material mixing in shock-driven turbulence, there are a wide variety of metrics used to measure mix growth. Inspired by reduced dimensionality RANS models, one commonly used metric is the width of the mixing layer or mix width. Here we use the product transverse averages of the mass fractions of air and SF$_6$ to define 
the mean mixing $\tilde{\phi}(x)$ as
\begin{equation}
\tilde{\phi}(x) = 4  \overline{Y_{\text{SF6}}} \; \left( 1 - \overline{Y_{\text{SF6}}} \right) .
\label{eq:MixWidth}
\end{equation}

We define the mix width $W$ as the length of the interval in the stream-wise direction where $\tilde{\phi}(x) \geq 0.75$.  While this definition is somewhat arbitrary, we have found that our results do not vary significantly when values of 0.7 or 0.8 are used. Our choice of 0.75 has been extensively used before \cite{Grinstein2011} and is designed to focus the analysis on the regions of strong mixing while also providing wide enough layers to limit realization noise for volume-averaged quantities.

Using this metric, we can define the volumes of primary mixing for both interfaces. We label the mixing layer associated with the first or left interface $L_1$ and the mixing layer associated with the second or right interface $L_2$. In Figure~\ref{fig:spacetime}, the two mixing layers are shown to evolve in width over time. We will make particularly detailed study of these two layers in subsequent analysis.
 
 Figure~\ref{fig:Vapor_4times} shows 3D volume renderings the SF$_6$ concentration at $t = 0.0$, 1.0, 2.0 and 3.0 ms for case S-32, highlighting the movement of the band as well as the growth of both interfaces. Of particular note is the transition from ballistic growth seen at $t = 2.0$ ms to the strong mixing seen by $t=3.0$ ms. While the quantitative details of these simulations differ, all follow the same basic evolution outlined here.  

\section{Resolution Study}

In this section, we compare cases H-32 and S-1, and V-1, H-1, and S-1, respectively, to address convergence issues. Firstly, we note that previous simulations using RAGE for a single planar-interface geometry have shown that large scale convergence to within chaotic variability can be achieved provided the initial interface perturbations are well resolved \cite{Grinstein2011}. Robustness of ILES macroscopic results is expected with sufficient resolution once there is sufficient separation in scales between the energy-carrying scales and the diffusive scales in the high Reynolds number limit \cite{Wachtor2013}. To investigate the degree to which our high-Re assumption is true, we have conducted several simulations at increased resolution.

 \begin{figure*}[t] 
 \subfigure[Reynolds Number]{\includegraphics[width=0.499\linewidth]{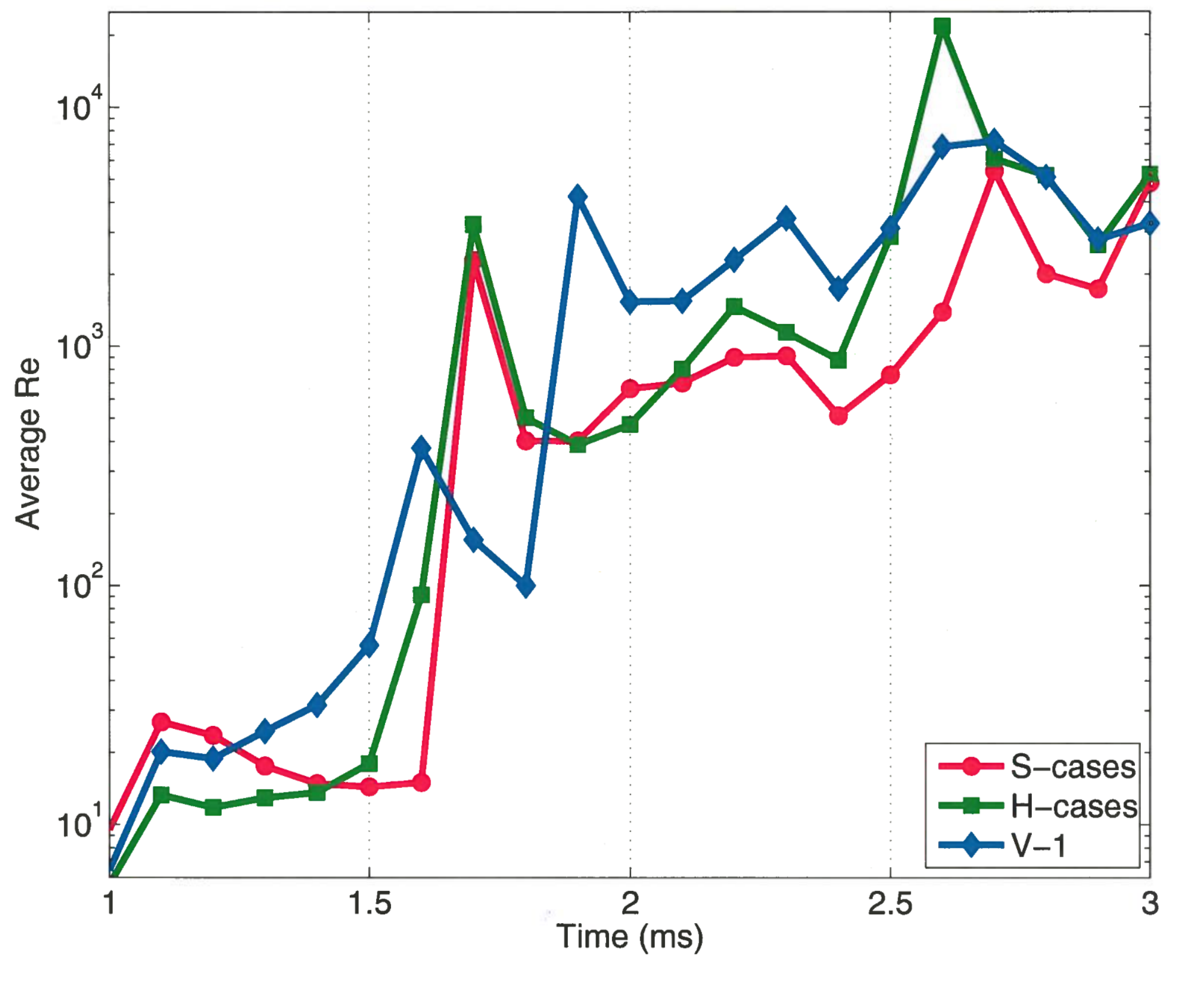}}
  \subfigure[Relative TKE]{\includegraphics[width=0.491\linewidth]{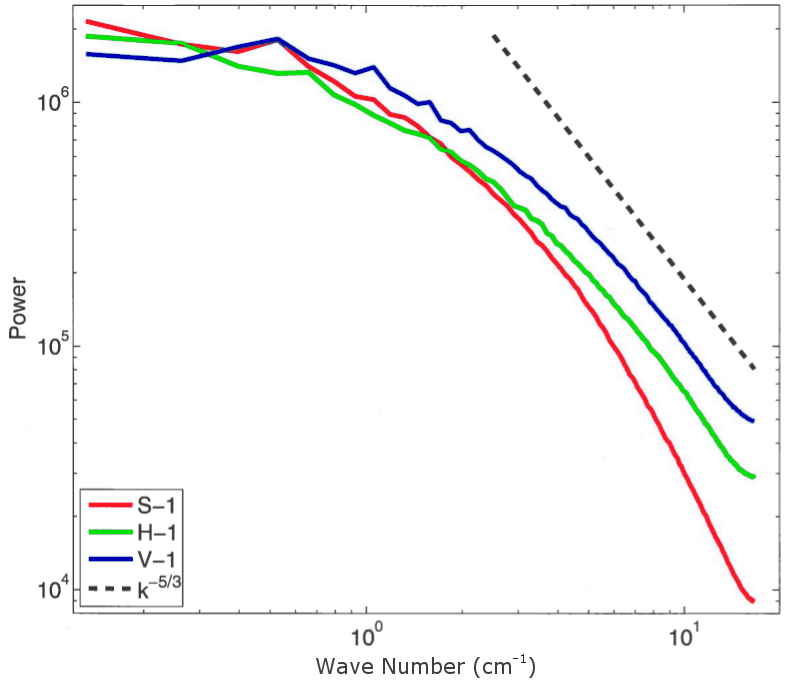}}
 \caption{(Color online) (a)~Ensemble-averged $\overline{\mathrm{Re}}$ for the S- and H-series cases, along with the $\overline{\mathrm{Re}}$ number for case V-1. The $\overline{\mathrm{Re}}$ for all three resolutions show significant variability in time, however there is a clear increase at late times for higher resolution simulations. (b)~Streamwise-averaged kinetic energy power spectra for cases S-1 (red/medium gray), H-1 (green/light gray), and V-1 (blue/ dark gray). All spectra are computed for wave numbers less than 16.7 cm$^{-1}$. All cases show similar spectral features at low wave numbers, but increasing resolution leads to extend inertial ranges. \label{fig:Re}}
 \end{figure*}

Secondly, it is possible that the effects of varying the initial spectral density may change when the dissipation of the system is changed. In ILES codes such as RAGE the effective (numerical) diffusion depends on the grid resolution.\cite{Wachtor2013, Zhou2014}  Thus our resolution study is actually a study of both increased resolution and decreased effective diffusion. In a well-developed turbulent system the dissipation seen on moderate scales should become independent of the scale at which dissipation occurs, as the cascade reaches a steady-state of transfer. However, for transitional flows this is less certain.  In running cases V-1, H-1, and S-1, and again in cases H-32 and S-32, identical initial conditions were used for both interfaces. Unlike the independent realizations of our initial conditions that were run for cases S-1, S-4, and S-32 which used unique random spectra for each interface in each realization, here we have used identical modes with identical phases in order to isolate the effects of resolution in our simulations. Ideally it would be desirable to also run multiple realizations (with randomly varying mode weights) at each resolution, however the computational demands of H- and V-series cases would make such variability studies impractical within our present scope; however expected sensitivity of results to such variations is demonstrated further below.

\subsection{Reynolds Number and Spectral Diagnostics}

When studying the resolution dependence of our results, we chose to first examine the quantities we would expect to vary in our models, namely Re and the kinetic energy spectrum. Only estimates for Re for these simulations are possible as our ILES framework does not include physical dissipation. Furthermore any discussion of Re must account for the transitional nature of the flows as the RMI develops from ballistic growth to mode interactions to turbulence and finally to decaying turbulence.  We can estimate an ILES effective Re  by finding a measure of the effective viscosity $\nu_\mathrm{eft}$ associated with resolution estimated as ratio of computed dissipation and squared strain-rate magnitude
using the raw simulated flow velocity \cite{Wachtor2013} -- distinct from the grid-scale viscosity associated with the numerical scheme specifics.   Zhou et al.~\cite{Zhou2014} have explored several methods for estimating the $\nu_\mathrm{eff}$. Here we will utilize two methods, which both posit that the effective viscosity can be represented by 
\begin{equation}
\nu_\mathrm{eff} = \frac{ \epsilon }{ \Omega } ,
\end{equation}
where $\epsilon$ is the energy dissipation rate and $\Omega = 1/2 ( \nabla \times \vec{v})^2$ is the enstrophy. The first method, which we will use to calculate what we term Re$_1$ uses the resolved energy flux spectrum 
\begin{equation}
\nu_\mathrm{eff} = \frac{ \epsilon }{ \Omega } \approx \frac{ - \int^{k^*}_0 \frac{ \partial E (k, t) }{ \partial t } \; dk }{ \Omega } ,
\end{equation}
where  $k^*$ is the wave number chosen at which the energy transfer rate has achieved a constant value, and $E(k,t)$ is the Fourier transform of the specific kinetic energy \cite{Zhou2014}. Using this prescription, we can then compute Re$_1$ as
\begin{equation}
\mathrm{Re}_1 = - \frac{U L \Omega }{\int^{k^*}_0 \frac{ \partial E (k, t) }{ \partial t } \; dk }
\end{equation}
where $U$ is the volume-averaged magnitude of the fluctuating velocity over the mixing layer, $L$ is the width of the mixing layer, and $\Omega$ is the volume-averaged enstrophy.

  \begin{figure*}[t]
 \subfigure[TKE]{\includegraphics[width=0.48\linewidth]{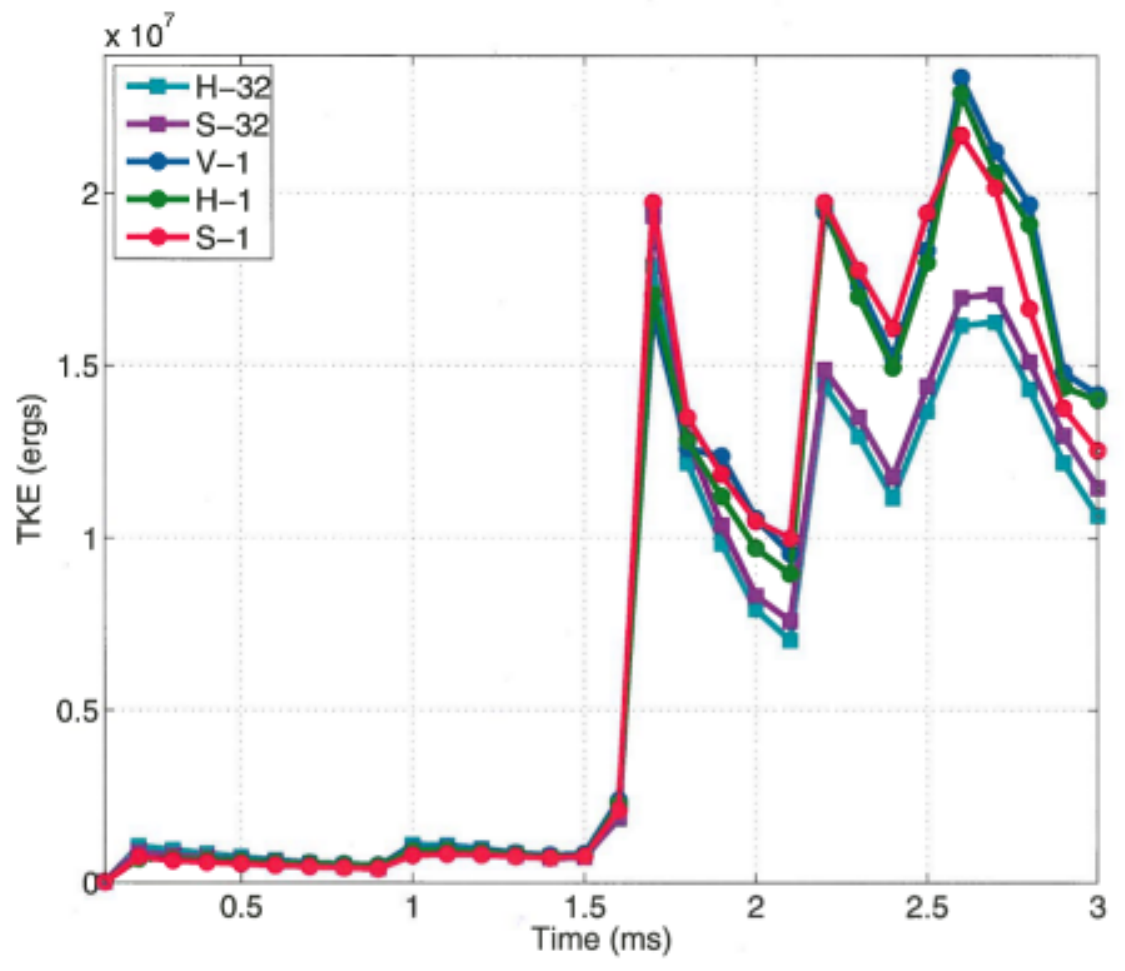}}
  \subfigure[TMX]{\includegraphics[width=0.48\linewidth]{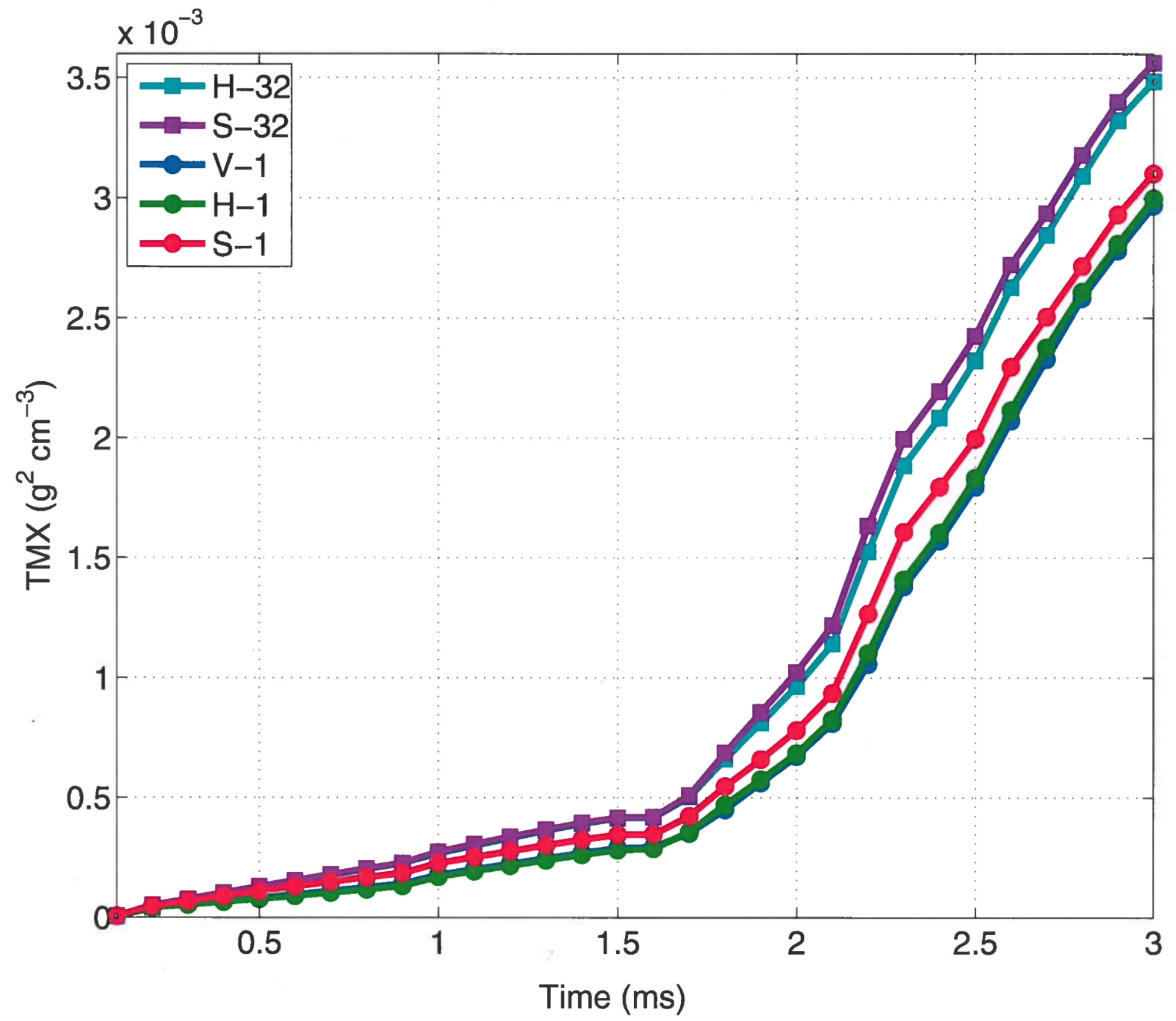}}
 \caption{(Color online) (a)~Turbulent kinetic energy (TKE) as a function of time for simulations S-32, S-1, H-32, H-1, and V-1. (b)~Total mixednes (TMX) for the same simulations. Note that lines for all five cases in both plots are present at all times, though they are often over-plotted by another case as values are very similar. For both quantities the evolution does not show significant changes as the numerical resolution is increased and the variation caused by changing the initial spectral density is considerably larger than the variation due to the change in resolution. In particular, cases H-1 and V-1 show less than 2.8\% variation in TKE and less than 1.1\% variation in TMX at late times. \label{fig:ResStudy}}
 \end{figure*}

The second method explored by Zhou et al. \cite{Zhou2014} uses the dimensionless parameter $D$, which is given by
\begin{equation}
D = \frac{ \epsilon L }{ U^3 }, 
\end{equation}
thus providing an alternative means of computing the dissipation rate $\epsilon$ in terms of the characteristic velocity $U$ and length $L$ scales if $D$ is known \cite{Zhou2014}. DNS models of forced and decaying turbulence have shown that $D$ asymptotically tends to a constant value of order unity with increasing Re \cite{Zhou2014}.  Assuming our Re is high enough that our simulations fall in the asymptotic regime, we can compute another estimate for the Re as
\begin{equation}
\mathrm{Re}_2 =  \frac{ L^2 \Omega } { U^2 }.
\end{equation}
Here, we have again used the mix width for our length scale $L$, the volume-averaged magnitude of the velocity fluctuations as our velocity scale $U$, and the volume-averaged enstrophy $\Omega$.

Interestingly, we find that both estimates provide very similar values for each simulation as a function of time. We compute what we term a mean Reynolds number $\overline{\mathrm{Re}}$ as 
\begin{equation}
\overline{\mathrm{Re}} = \frac{1}{2} \left( \mathrm{Re}_1 + \mathrm{Re}_2 \right) .
\end{equation}
We further compute ensemble means for all cases at a given resolution at each time in order to provide an additional reduction in noise. Figure~\ref{fig:Re}(a) shows the mean Re for each resolution as a function of time. While there is considerable noise in these calculations, there is also a clear trend for higher resolutions to produce larger Re at late times,  approaching the (integral-scale based Re) mixing transition threshold values between $10^4$ and $2 \times 10^4$ \cite{dimo}.  

%
% bibitem{dimo}
%P.~E.~Dimotakis.
%\newblock The mixing transition in turbulent flows.
%\newblock {\em J.~Fluid Mech.}, 409:69, 2000.
%

An additional diagnostic of our resolution study is examining the length of the simulated inertial range in the kinetic energy power spectra of our mixing layers. Higher Re flows should exhibit a greater separation in scales between the energy-containing large scales and the dissipation range, and we should find a larger self-similar inertial subrange with expected Kolmogorov scaling of kinetic energy power law $\sim k^{-5/3}$.  Figure~\ref{fig:Re}(b) shows the fluctuating kinetic energy power spectra for the back mixing layer at $t = 2.8$ ms. Case V-1 clearly shows the longest inertial range, while case H-1 and case S-1 suggest progressively smaller inertial ranges.

\subsection{Large-Scale Behaviors}

While Re and small-scale behaviors of our models clearly do vary with resolution, we will show that the S-series cases are sufficiently resolved in to achieve some level of convergence on the large-scale features of interest to us as we increase resolution in cases H-1, H-32, and V-1. Specifically, we will examine the volume-integrated turbulent kinetic energy and mixedness as a function of resolution.

Figure~\ref{fig:ResStudy}(a) shows the evolution of the turbulent kinetic energy TKE over the course of the simulation for five cases. We define TKE as 
\begin{equation}
\mathrm{TKE} = \int_{\mathcal{V}} \frac{\rho}{2} \left[ \left( v_x - \bar{v}_x \right)^2 + \left( v_y - \bar{v}_y \right)^2 + \left( v_z - \bar{v}_z \right)^2 \right] \; d \mathcal{V},
\label{eq:TKE}
\end{equation}
where $\mathcal{V}$ represent the entire simulated volume. TKE thus removes the kinetic energy associated with any flows along the axis of the shock tube, most importantly the shocks themselves. We have plotted TKE for cases V-1, H-32, H-1, S-32, and S-1. Cases H-32 and S-32 show very similar overall evolution with some deviation, as would be expected for chaotic systems like ours. Cases V-1, H-1, and S-1 also show very similar overall behavior. Careful examination does show that for both values of $N_{IC}$ the S-series resolution simulations both consistently show between 2\% and 5\% more mixing after reshock than their H-series counterparts, indicating that there may be some aspects of this problem that are not fully numerically converged, however when compared to the variation caused by changes to the initial spectral density the variations due to increased numerical resolution are between 3 and 6 times smaller. Case V-1 shows less than 3\% variation from case H-1 at all times and no clear trend relative to case H-1. We attribute this variation, which is largest after 1.6 ms, to chaotic variability.

 \begin{figure*}[t] 
 \subfigure[TKE]{\includegraphics[width=0.495\linewidth]{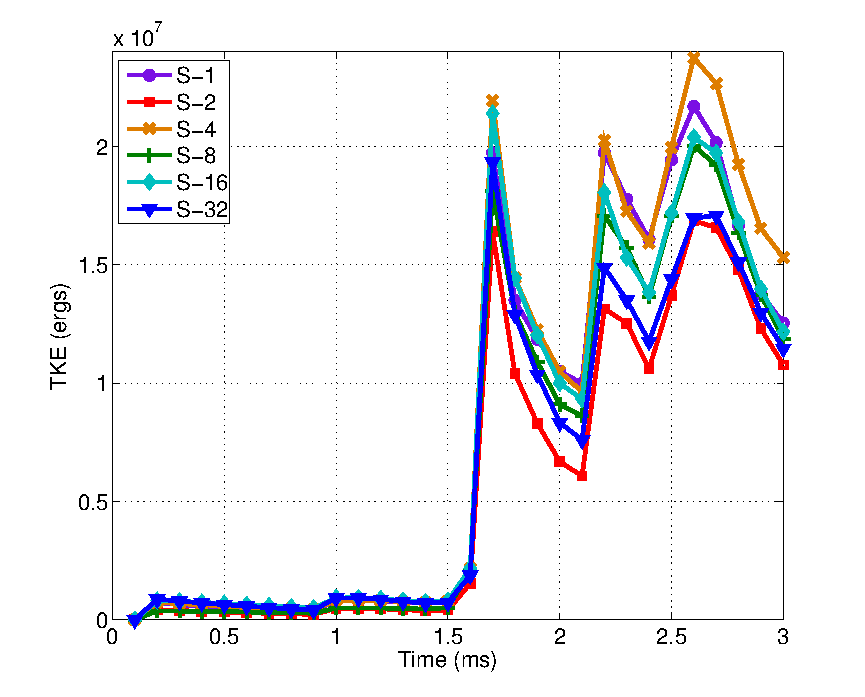}}
  \subfigure[Relative TKE]{\includegraphics[width=0.495\linewidth]{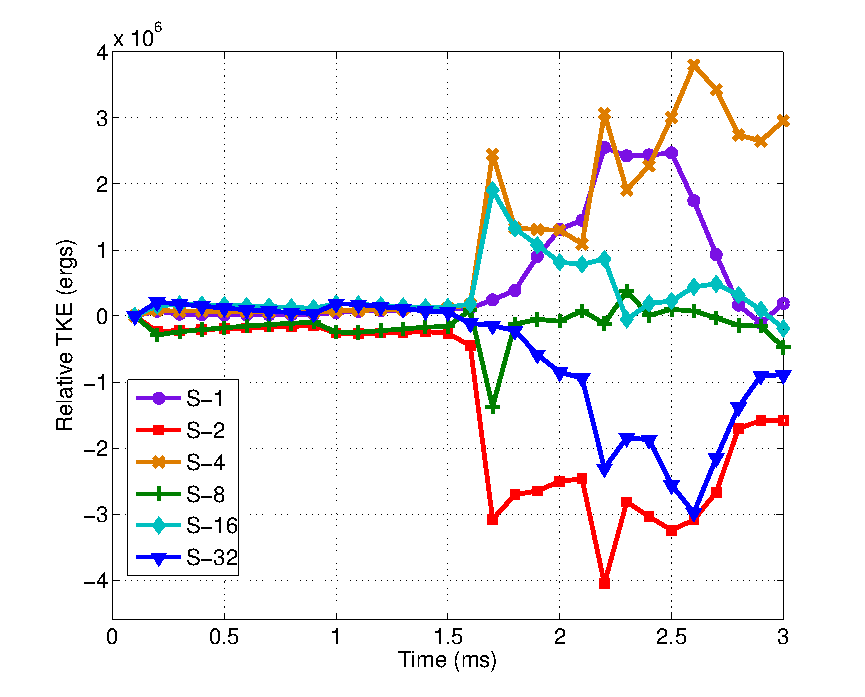}}
 \caption{(Color online) (a)~Turbulent kinetic energy as a function of time for simulations S-32 through S-1. (b)~Relative TKE for the same simulations where the ensemble mean at each time from (a) has been subtracted for each simulation. While strong variations are seen, there is no obvious trend with $N_{IC}$. \label{fig:TKE}}
 \end{figure*}

Figure~\ref{fig:ResStudy}(b) shows the evolution of the volume-integrated mixedness of the same five simulations. In our RAGE simulations, the composition inside a computational cell is assumed to be uniform throughout that cell, thus materials are numerically treated as if they are a uniform molecular mix. We term this mixedness.  One measure of mixedness is the volume-integrated, density-weighted mixedness TMX. Like TKE, TMX is a volume-integrated measure of the mixing between the air and SF6 in our domain. TMX is defined as 
\begin{equation}
\mathrm{TMX} = \int_{\mathcal{V}} \rho^2 Y_{\mathrm{SF6}} Y_{\mathrm{air}}  \; d \mathcal{V},
\label{eq:TMX}
\end{equation}
where $Y_\mathrm{air}$ can also be expressed as $ 1 - Y_\mathrm{SF6}$. Thus TMX is the product of the partial mass fractions for the two materials in a computational cell at any instant, integrated over the computational domain. We can consider the volume integrated TMX over the whole domain, thereby including both interfaces in the same metric, or we can divide the domain into two halves at the center of the SF$_6$ band, thereby allowing us to examine the mixedness for each interface separately. We also examine the normalized total mixedness, where the normalization is provided by the TMX computed for our IC.
 
In an ILES code such as RAGE, increases in mixedness are only facilitated by numerical diffusion of species concentration gradients at or near the grid scale. The numerical operators in RAGE are essentially inviscid in regions without small-scale gradients. In our IC, only cells on the interfaces at the front and back of the SF$_6$ band have any mixedness. All subsequent growth in TMX is thus a useful diagnostic of the effectiveness of the turbulent cascade in moving energy (and hence material interfaces) to sufficiently small scales where numerical dissipation can act. Generally this numerical diffusion is minimized in an ILES framework, but is still likely to be much greater than realistic molecular diffusion for this application.

   \begin{figure*}[t]
 \subfigure[Mix Width for $L_1$]{\includegraphics[width=0.495\linewidth]{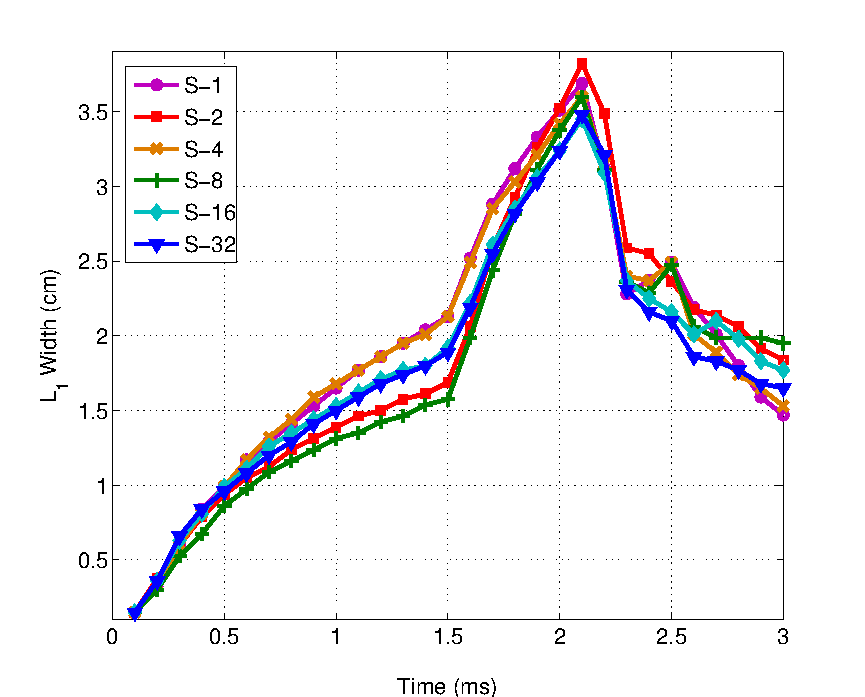}}
  \subfigure[Mix Width for $L_2$]{\includegraphics[width=0.495\linewidth]{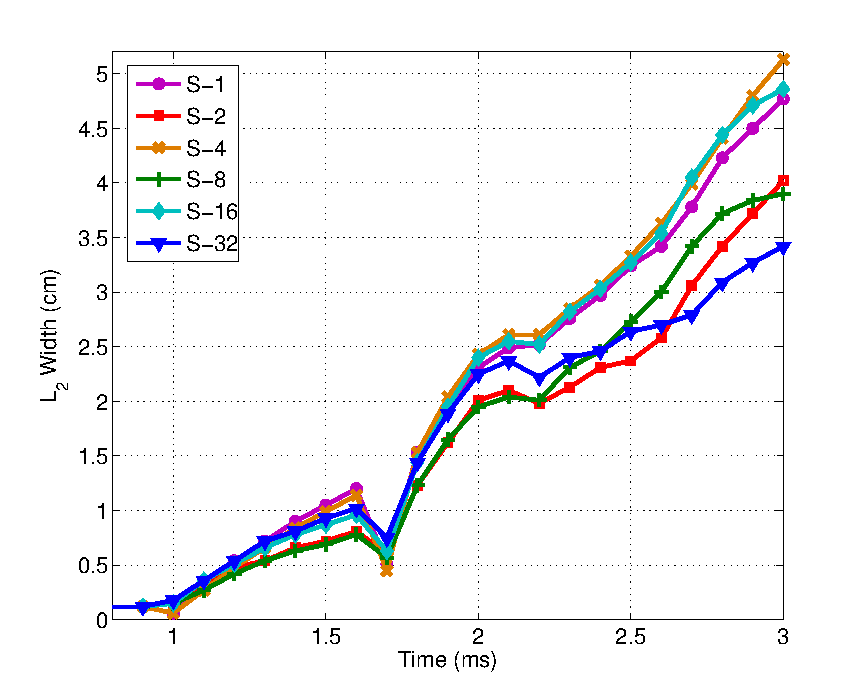}}
   \subfigure[Relative Mix Width for $L_1$]{\includegraphics[width=0.495\linewidth]{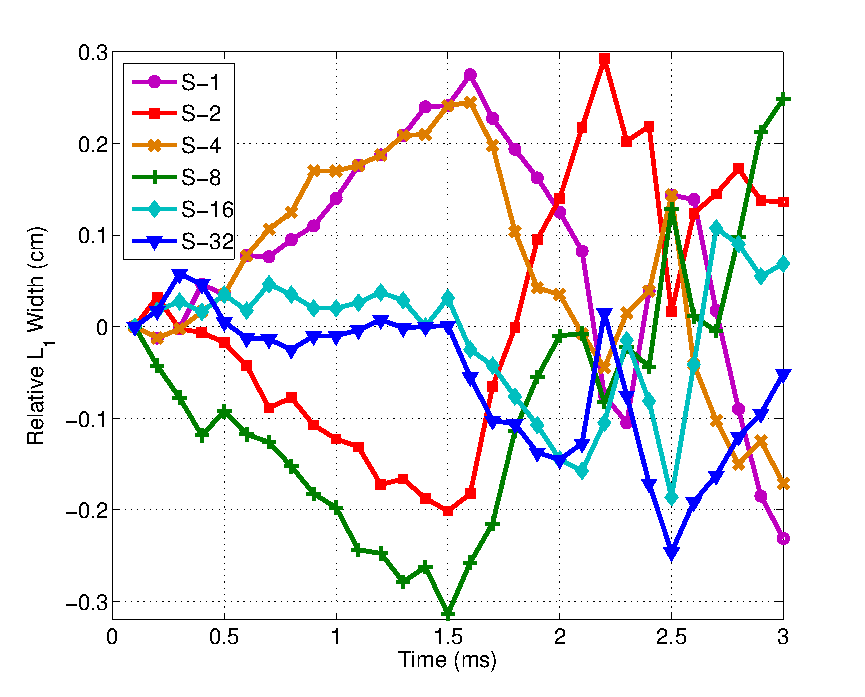}}
  \subfigure[Relative Mix Width for $L_2$]{\includegraphics[width=0.495\linewidth]{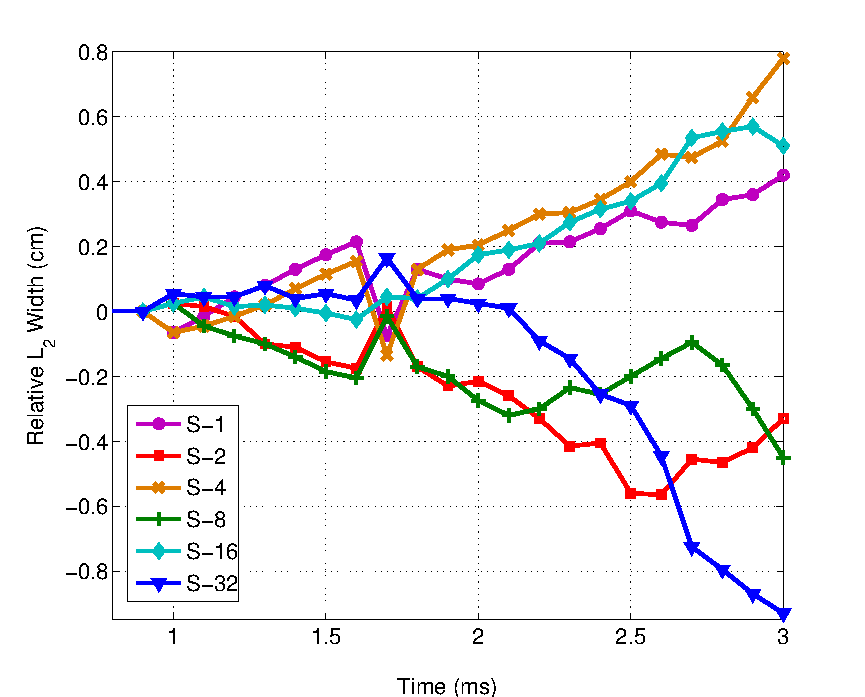}}
 \caption{(Color online) (a)~Time-evolution of the width of the mixing layer $L_1$ for cases S-1 through S-32. The mix widths grow slowly until the rarefaction wave hits it at about $t=1.6$ ms, after which all six cases show a rapid expansion of the layer. This expansion is reversed upon reshock, which recompresses the the layer. (b)~Time-evolution of the mix width for the second interface $L_2$ for all six cases. The mix width grows slowly until reshock at about $t = 1.6$ ms where a sharp compression occurs and is followed by continued expansion. (c)~Relative mix width for the front mixing layer $L_1$ where the ensemble mean of the six cases at each time has been subtracted. No clear trends with $N_{IC}$ can be seen either in the orderly evolution prior to reshock or the more chaotic behavior after reshock. (d)~Relative mix width for the back interface $L_2$ where the ensemble mean of the six cases has again been subtracted at each time. While the mix widths are less chaotic than in $L_1$, there is no trend with $N_{IC}$. \label{fig:MixWidth}}
 \end{figure*}

As with TKE, Figure~\ref{fig:ResStudy}(b) demonstrates that TMX for all five cases follow the same general trend with slow growth prior to $t = 1.6$ ms followed by much more rapid mixing. Cases H-32 and S-32 show very similar trends with deviations of less than 4\% over their entire evolution. Cases H-1 and S-1 also show highly similar evolution with maximum deviation of only 6\%, while cases H-32 compared with H-1, and S-32 compared with S-1 show much greater disagreement of as much as 18\% and 16\%, respectively. In the total mixedness, the variation between cases H-1 and V-1 is very small. At $t = 2.0$ ms case V-1 shows 0.73\% less mixedness than case H-1, while at $t = 3.0$ ms case V-1's TMX is 1.12\% smaller than case H-1's. Unlike TKE, TMX does show a very small but systematic decrease between cases H-1 and V-1. We speculate that continued increases in resolution may yield asymptotically smaller reductions in TMX. Thus for the purposes of examining the impact of initial spectral density on mixing, we feel confident that our numerical resolution is sufficient.

For both global measures, increased resolution has smaller changes to the volume-integrated quantities measuring turbulent kinetic energy and mixedness. Thus we believe that the primary effect in our models is the change in initial spectral density rather than any issues arising from insufficient numerical resolution. 
 
 \section{Effects of Spectral Density of Initial Perturbations} 
 
   \begin{figure*}[t]
 \subfigure[TMX]{\includegraphics[width=0.495\linewidth]{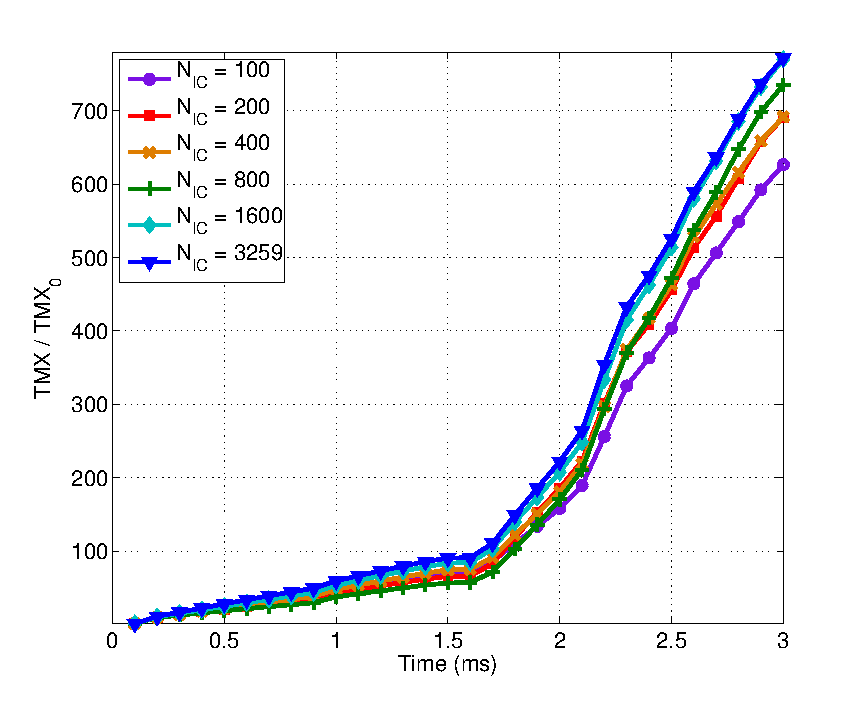}}
  \subfigure[Relative TMX]{\includegraphics[width=0.495\linewidth]{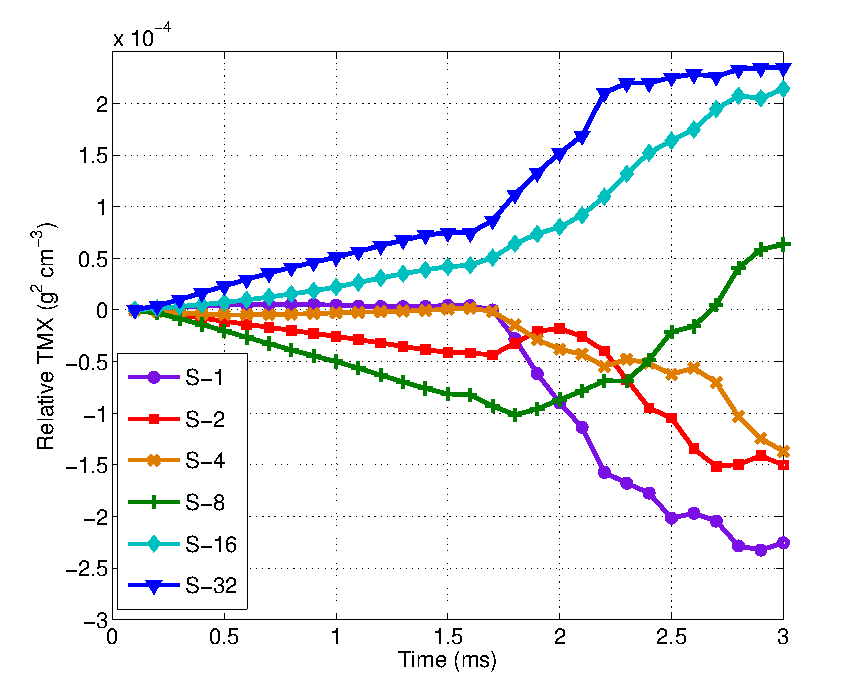}}
   \subfigure[TMX by Interface]{\includegraphics[width=0.495\linewidth]{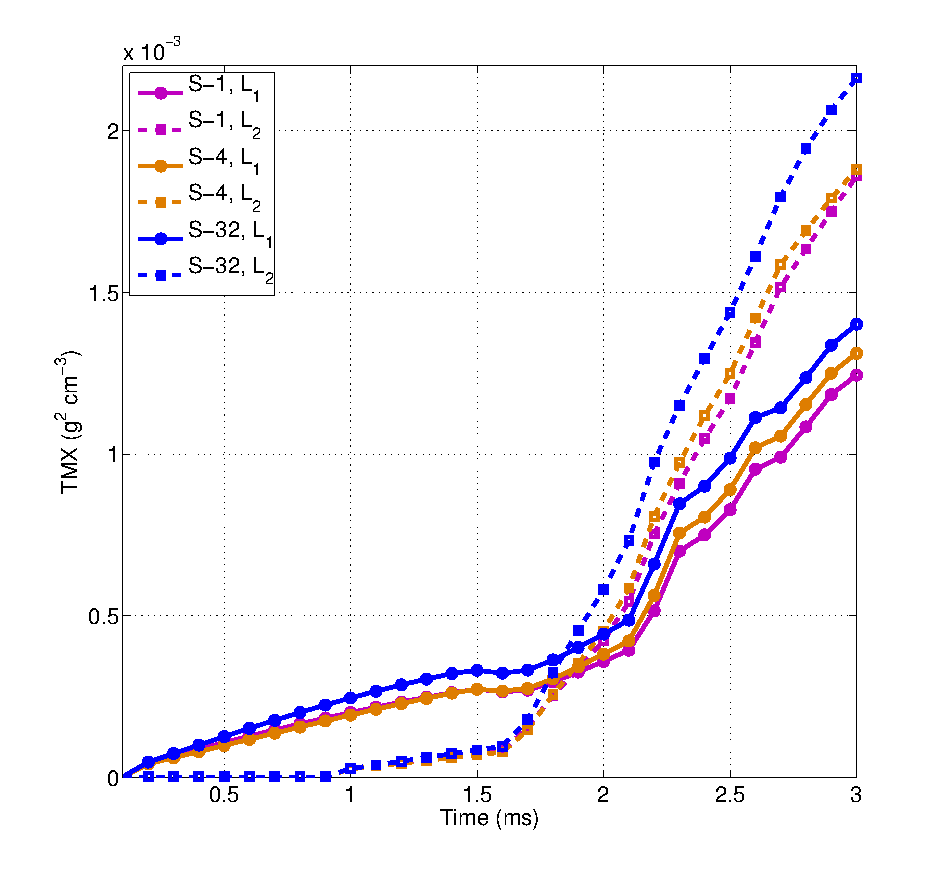}} 
 \caption{(Color online) (a) Volume-integrated mixedness TMX for cases S-32 through S-1 as a function of simulation time. Values have been normalized by the values of TMX for each case just prior to the first impact of the shock on the front interface, which we term TMX$_0$. Two clear phases of mixing can be seen with mixing before 1.5 ms showing linear behavior followed by much more rapid, turbulent mixing. (b)~Relative TMX for the same simulations with the ensemble mean subtracted at each time. (c)~TMX for each interface for cases S-1, S-4, and S-32. While the initial linear phase shows little dependence on $N_{IC}$, the later turbulent phase reveals a clear trend where cases with greater $N_{IC}$ exhibit more mixing. By $t = 2.5$ ms there is a monotonic relation between $N_{IC}$ and TMX for both interfaces individual and over the entire domain. \label{fig:TMX}}
 \end{figure*}
 
   \begin{figure}
 \includegraphics[width=\linewidth]{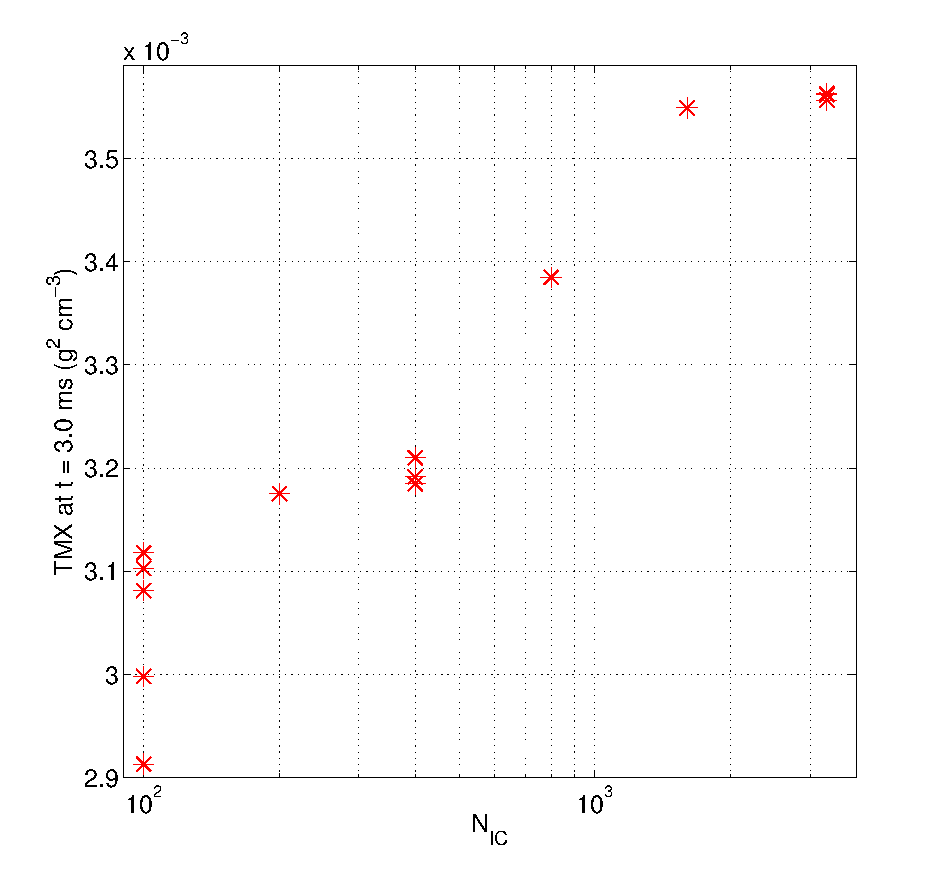}
 \caption{(Color online) Volume-integrated mixedness at $t = 3.0$ ms for six values of $N_{IC}$. Unique randomly generated instances of the initial perturbation spectra were used for each simulation, with five realizations of case S-1 and three each for cases S-4 and S-32. For case S-32, all three symbols are plotted here though they are visually difficult to distinguish. While there is some evidence for significant variability due to realization noise, particularly for small values of $N_{IC}$, the larger trend of less mixedness with decreasing $N_{IC}$ is clear. \label{fig:RealNoise}}
 \end{figure}

The primary purpose of this paper is to investigate the effects of the spectral content of initial interface perturbations on the growth of the RMI, the coupling of modes through nonlinear behaviors, and the development of turbulence. To that end, we have conducted a series of otherwise identical simulations with variable numbers of Fourier modes in their initial interface perturbations. These simulations are labeled S-1, S-2, S-4, S-8, S-16, and S-32 (see Table~\ref{table:Cases}). All six cases follow the same general time evolution as shown for case S-32 in Figure~\ref{fig:spacetime}, however significant variations are seen in the properties of the resulting mixing layers around both interfaces both before and after reshock.

Figure~\ref{fig:TKE}(a) shows the evolution of the total turbulent kinetic energy (TKE) in the computational domain. Examining the time-evolution of TKE, we can see sharp increases at 0.1 ms when the shock first impacts $L_1$, at 0.9 ms when the main shock impacts $L_2$, around 1.6 ms when the rarefaction wave hits $L_1$, and about 1.6 and 2.2 ms when the reflected shock excites additional motion at each interface. All six cases follow the same general timing, however the relative levels of TKE achieved differ widely.

Figure~\ref{fig:TKE}(b) shows evolution of TKE in all six cases with the ensemble mean at each time subtracted. This highlights the variation is the six cases. Variations of as much as 50\% can be seen between cases, most notably between cases S-4 in brown and S-2 in red. This is striking as all cases have IC with equal spectral slopes and surface standard deviations. The only difference here is the number of modes applied as interface perturbations. There does not, however, appear to be a systematic trend with $N_{IC}$ at either early or late times.

As discussed above,  the mix width $W$ is defined as the length of the interval in the stream-wise direction where $\tilde{\phi}(x) \geq 0.75$.  Figure~\ref{fig:MixWidth} shows the widths of the mixing layers at both the front and back of the SF$_6$ band as a function of simulation time. Figure~\ref{fig:MixWidth}(a) shows the extent of $L_1$, while panel (b) shows the width of $L_2$. For both mix widths, we see slow growth after the primary shock. For $L_1$ we see a sharp increase in the mix width upon rarefaction and then a sharp decrease as reshock compresses the mixing layer. For $L_2$ we see a brief compression of the mixing layer on reshock followed by continued expansion of the mixing layer. All six cases show similar behaviors on both interfaces.

     \begin{figure*}
 \subfigure[$Y_{SF6}$ for Case S-32]{\includegraphics[width=0.495\linewidth]{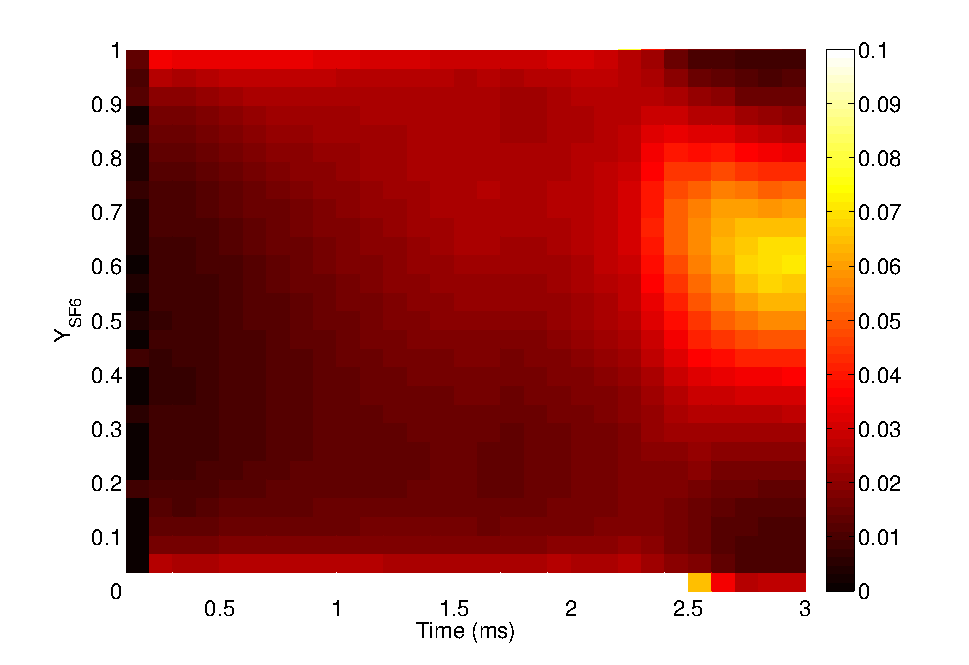}}
  \subfigure[$Y_{SF6}$ for Case S-1]{\includegraphics[width=0.495\linewidth]{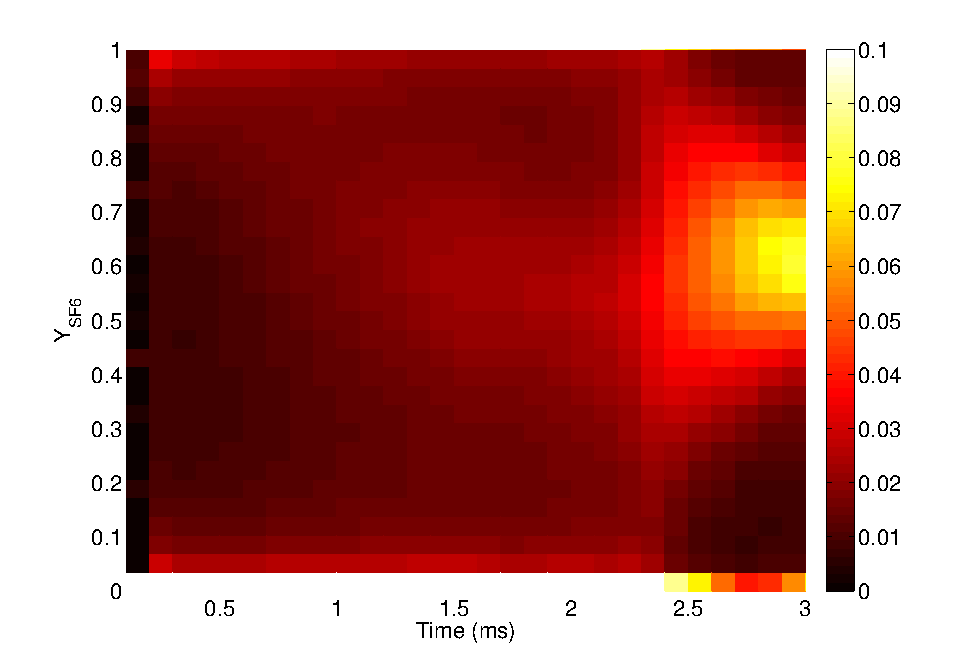}}
    \subfigure[Ratio of Case S-1 to S-32]{\includegraphics[width=0.495\linewidth]{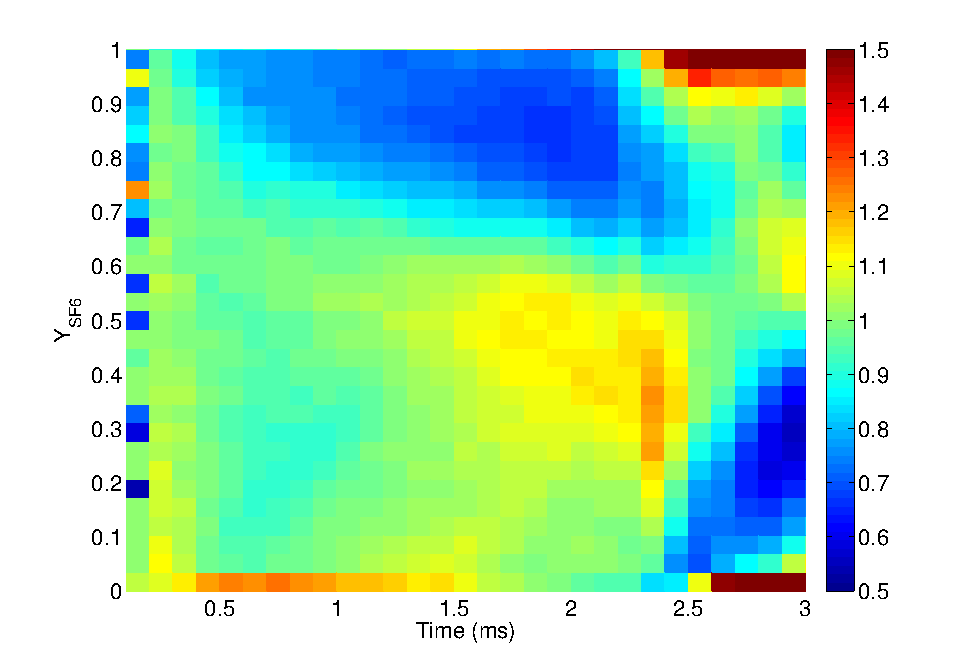}}
 \caption{(Color online) Time-evolution of the normalized probability distribution for $Y_{SF6}$ in mixing layer $L_1$ for (a)~Case S-32 and (b)~Case S-1 with light tones indicating larger fractions of the mixing layer at a given concentration and dark tones showing smaller fractions of the mixing layer volume.  (c)~The ratio of (b) to (a), with blue to green (darker gray) tones indicating larger values for case S-32 and red to yellow (lighter gray) tones indicating larger values for case S-1. \label{fig:SF6_PDF}}
 \end{figure*}

Figure~\ref{fig:MixWidth}(c) shows the relative mix width for $L_1$, while panel (d) shows the same for $L_2$. To construct the relative mix width we subtract the ensemble mean for all six cases at each time from each curve, thus removing the general trend to focus on the variations between cases. Figure~\ref{fig:MixWidth}(c) shows that while there are significant variations in the mix widths of as much as 20\% prior to rarefaction, there are no clear trends as the initial spectral density is changed. Particularly after rarefaction and reshock, the mix width of $L_1$ shows dramatic and chaotic variability between the six cases. The relative mix widths for $L_2$ are somewhat less chaotic, but still no trend is seen with initial spectral density.

Our turbulent mixing measures are usefully characterized by the two different mixing processes captured with ILES. 
Mix width is largely due to large-scale entrainment promoted by bubbles and spikes, while mixedness is driven by stirring associated with smaller scale velocity gradient fluctuations.  In Figure~\ref{fig:TMX}(a) we can see that, much as with TKE, the time evolution of TMX follows the same general trend 
for all six cases. There is a slowly-growing phase up until about 1.6 ms when the first rarefaction wave hits the front mixing layer leading to the 
large jump in TKE seen in Figure~\ref{fig:TKE}(a) at that time. After this point TMX grows much more rapidly for all cases.

The variations in TMX between the six cases are highlighted in Figure~\ref{fig:TMX}(b) where we have removed the ensemble mean at each time, thus eliminating the general trend in TMX with time. There is a slow divergence between the six cases until $t = 1.6$ ms when reshock of $L_2$ and rarefaction of $L_1$ occur. After $t = 1.6$ ms the divergence between the cases accelerates. Variations between cases of as much as $\pm 30$\% from the ensemble mean are seen even at early times. TMX shows a strong dependence on $N_{IC}$ at late times. By $t = 3.0$ ms, the relation between $N_{IC}$ and TMX is monotonic. The more dense the spectrum of initial perturbations applied to the interfaces, the more mixing at late times. Figure~\ref{fig:TMX}(c) shows that this trend holds for both interfaces individually. The mixedness of both the front and back of the band see little or no dependence on $N_{IC}$ until $t = 16.0$ ms when the rarefaction wave hits the front interface and the back interface is reshocked. Mixing at the front interface is again accelerated when it too is reshocked.

\begin{figure*}[t]
 \subfigure[]{\includegraphics[width=0.495\linewidth]{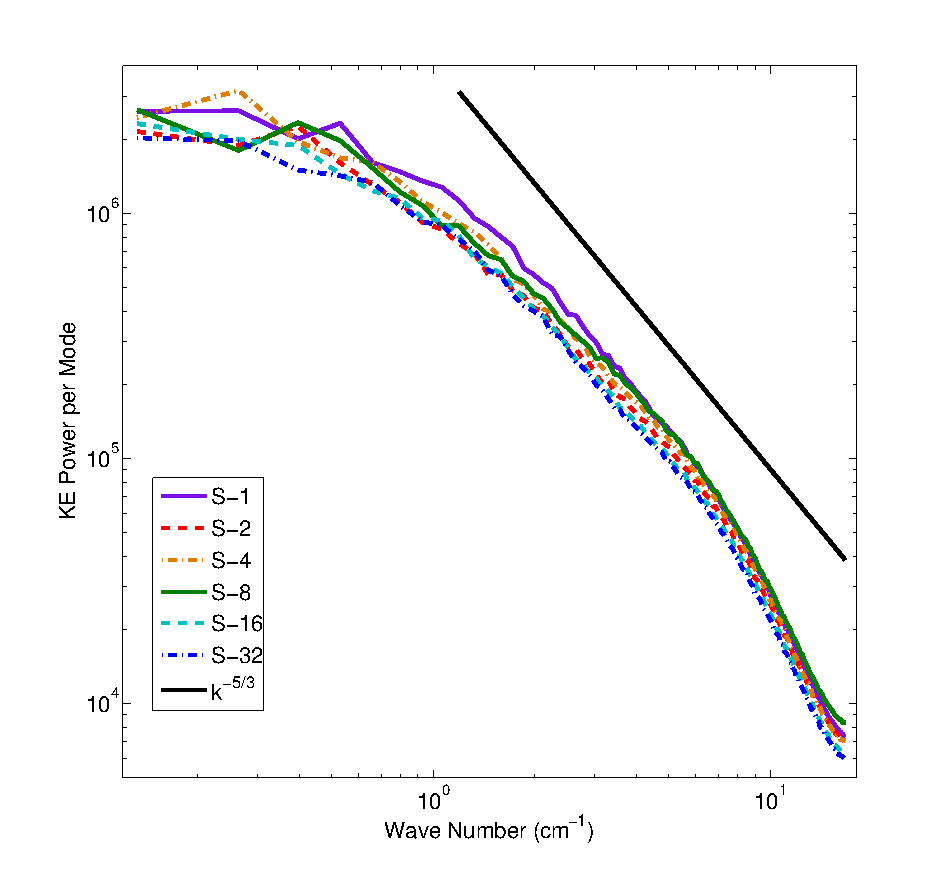}}
  \subfigure[]{\includegraphics[width=0.495\linewidth]{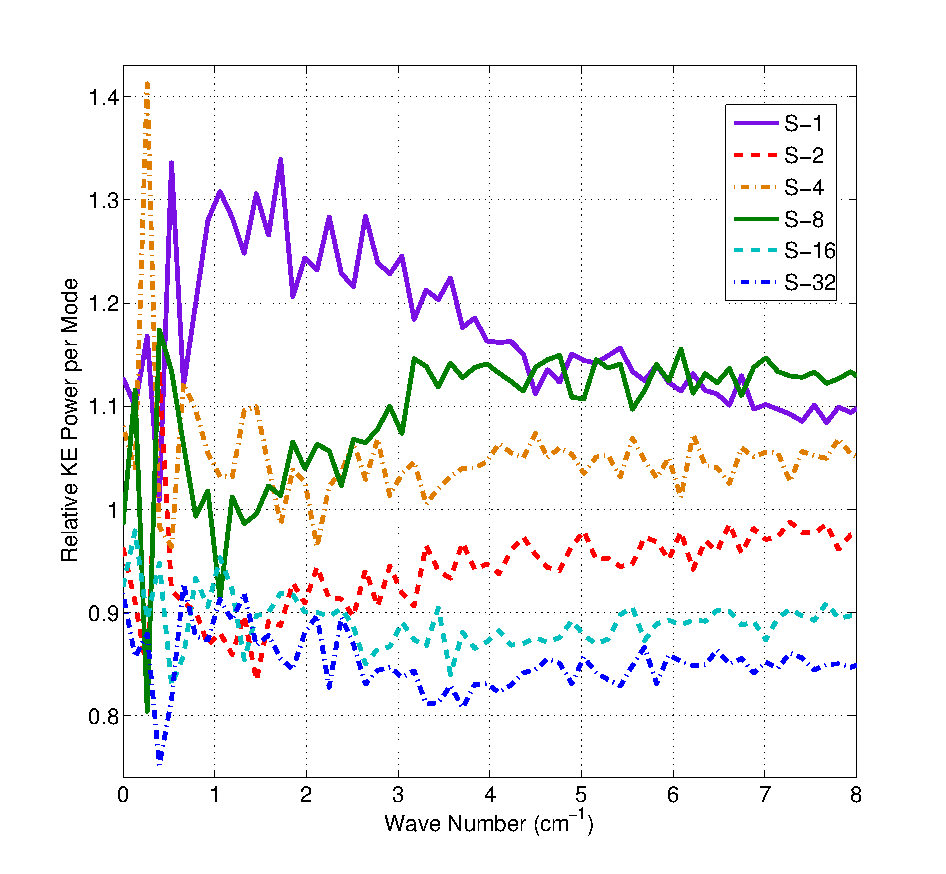}}
 \caption{(Color online) (a) Streamwise-averaged transverse power spectra of kinetic energy in mixing layer $L_1$ at $t = 3.0$ ms for cases S-32 through S-1. All cases show a turbulent power spectra with a small but significant inertial range. (b)~Relative power spectra of kinetic energy as a fraction of the ensemble mean for all six cases. While low $k$ power is highly variable and shows no clear dependence on $N_{IC}$, the high wave number portion of the spectra shows a clear dependence on $N_{IC}$.  \label{fig:SpecKE}}
 \end{figure*}

Case S-8 is particularly interesting as it shows the least TMX of all six simulations at $t = 1.6$ ms but rapidly increases to pass cases S-1, S-2, and S-4 by $t = 2.4$ ms. This indicates that while mixing in the early growth of the interface perturbations due to the RMI might not be as strongly dependent on the spectral density of the initial perturbations as the late stages after rarefaction and reshock when the flows become much more turbulent.

The dependence of TMX at late times on $N_{IC}$ shown in Figure~\ref{fig:TMX} naturally brings up the question of realization noise due to the random selection of modes and their phases for our IC. To help address this issue we have run additional random realizations of cases S-32, S-4, and S-1. Figure~\ref{fig:RealNoise} shows TMX at $t = 3.0$ ms for the S-series of simulations, including five unique, randomly generated realizations of case S-1 and three realizations each for cases S-4 and S-32. As a liberal estimate of the realization noise for each value of $N_{IC}$, we take the largest difference between TMX fora given value of $N_{IC}$ and divide by the average over all realizations. This yields realization noise levels of 0.2\% for case S-32, 0.8\% for case S-4, and 6.7\% for case S-1. It is not surprising that realization noise scales inversely with $N_{IC}$ as the cases with fewer modes are more susceptible to changes in any single mode. This suggests that our reported relationship between initial spectral density and late-time mixedness is not simply an artifact of particular realizations of our random initial perturbations and phases.
 
Another diagnostic of the mixing present in these simulation is the normalized probability distribution function (PDF) of the mass fraction of SF$_6$ in the mixing layers at the front and back of the band. We define these mixing layers using the mixing parameter $\phi$ defined in Equation~\ref{eq:phi}. We can further examine the evolution of the PDF for each simulation over time. Figure~\ref{fig:SF6_PDF} shows the PDFs for $Y_\mathrm{SF6}$ over the evolution of cases S-1 and S-32. For simplicity we have chosen to present only the two limiting cases, but the intermediate simulations show behaviors which both qualitatively and quantitatively fall between there two cases. Figure~\ref{fig:SF6_PDF}(a) shows the time evolution of the PDF for $Y_\mathrm{SF6}$ for case S-32 with 30 bins at 0.1 ms intervals. Lighter colors indicate larger fractions of the mixing layer $L_1$. Large fractions of the mixing layer are dominated by either pure SF$_6$ or pure air until 2.5 ms when the most of the mixing layer rapidly becomes mixed with $Y_\mathrm{SF6}$ between 0.4 and 0.8. Figure~\ref{fig:SF6_PDF}(b) reveals qualitatively similar behavior for case S-1. 

Figure~\ref{fig:SF6_PDF}(c) shows the ratio of the PDFs over time in Case S-1 to those in Case S-32, which blue tones indicating larger values for a given bin at a given time in Case S-1 and red values showing larger values for Case S-32. Prior to 2.5 ms, the primary difference between these two simulations is that Case S-1 shows somewhat more volume with $Y_\mathrm{SF6} \lesssim 0.6$ while Case S-32 shows more volume for $Y_\mathrm{SF6} \gtrsim 0.6$. After 2.5 ms, however, sharp differences become more apparent. Case S-1 shows a strong excess of volume with pure or nearly-pure materials compared to Case S-32, with more than 50\% larger fractions of the mixing layer in the lowest and highest bins. Case S-32 shows as much as 50\% larger fractions of the mixing layer with $0.1 \lesssim Y_\mathrm{SF6} \lesssim 0.4$ and roughly equal values for $Y_\mathrm{SF6} \gtrsim 0.5$. From this we can see that Case S-1 includes more ``blob''-like behavior of the two fluids as resolved nearly-pure chunks of both gases pass each other, while Case S-32 shows much more mixing of the two gases at the grid-scale.

Figure~\ref{fig:SF6_PDF} shows that increased initial spectral density leads to enhanced mixing at the grid-scale. This may be due to the inhibition of a turbulent cascade of energy to small scales in cases where the initial spectral density is low. In effect, the enhanced mixing may be the result of the IC in case S-32 more closely approximating the cascade established naturally by turbulence. To test this concept, we need to explore the spectral content of the irregularities in the mixing layers.

We construct the streamwise-averaged transverse kinetic energy power spectra by first isolating the volumes for each mixing layer, then taking the 2D Fourier transform of the kinetic energy field in the transverse directions at each layer in the stream wise direction. We then sum the 2D spectra along arcs of constant wave number magnitude $k_m$, which we define as
\begin{equation}
k_m = \sqrt{k_y^2 + k_z^2} .
\end{equation}
In this way our transverse spectra for each slice in the streamwise direction is designed to avoid preferring either the $y$ or $z$ direction. Finally, we average the transverse spectra in the streamwise direction over each mixing layer. Ideally we would like to take a 3D power spectra in order to both study the anisotropy of the transitional turbulence, but for these simulations there is insufficient resolution in the streamwise direction for such a procedure.

Figure~\ref{fig:SpecKE}(a) shows the streamwise-averaged transverse kinetic energy power spectra for $L_1$ at $t = 3.0$ ms in cases S-1 through S-32. All six cases show similar power spectra with energy primarily in the large-scales with $k_m \lesssim 2$ cm$^{-1}$. Scales with $ 2 \lesssim k \lesssim 6$ cm$^{-1}$ are consistent with the establishment of a rough inertial range of turbulence and display an approximate scaling as $k_m^{-5/3}$. The effects of numerical dissipation are clearly evident on scales with $k \gtrsim 8$ cm$^{-1}$.

To highlight the variation between cases, we compute the ensemble mean spectra by averaging the power spectra over all six cases at each wave number and then dividing the power spectra for each case by the ensemble mean power spectra. Figure~\ref{fig:SpecKE}(b) shows the fraction of kinetic energy power per mode for each case relative to the ensemble mean. We have purposefully omitted modes with $k_m > 8$ cm$^{-1}$ to focus on the energy-containing and inertial scales of the flow. The energy-containing scales are quite noisy and do not reveal any clear dependence on the initial spectral density of the simulation, however the inertial scales do. For $ 2 < k < 6$ cm$^{-1}$ the power per mode is an almost monotonic function of the initial spectral density. Cases with lower initial spectral density show generally more power in their inertial scales, indicating that power has been less efficiently moved to small scales where it can then be dissipated.  

 \section{Summary and Conclusions}
 
 In this paper we have presented a series of numerical experiments investigating shock-driven turbulent mixing due to the Richtmyer-Meshkov instability. Our models include both rarefaction and reshock, which trigger transition to turbulent behavior. Specifically, we have investigated the effect of the initial spectral density of the perturbations, or the number of Fourier modes over a given wave number band, applied to the two material interfaces as IC. We have shown that increasing the initial spectral density leads to enhanced mixing, especially after reshock when the mixing layers show more well-developed turbulence. We have also shown that evidence from our simulations indicates that the primary cause of this variability in mixing is the inhibition or promotion of the turbulent cascade. When the initial spectrum of perturbations is sparse additional time is required for the turbulent cascade to populate the Fourier modes needed for the establishment of an inertial range, while a dense initial spectrum of interface perturbations is much more readily adapted to provide the needed turbulent cascade to small scales where motions and material can be diffused by our numerical treatment.
 
 We have also investigated the dependence of initial spectral density on the width of the mixing layers at both the front and back of our high-density band. In both cases no clear dependence on initial spectral density is observed. This indicates that the initial spectral density has little impact on the formation and growth of bubbles and spikes for broadband initial perturbations.
 
More broadly, this work reinforces the growing understanding of the importance of IC in determining the late-time mixing properties of  RMI-driven turbulence. Previous theoretical, computational, and experimental efforts have explored the impact of  single and multi-mode characteristics, spectral slope, and amplitude of initial perturbations \cite[e.g.,][]{Banerjee2009, Thornber2010, Gowardhan2011, Balasubramanian2012, Jacobs2013}. To this we presently add initial spectral density as yet another parameter which shows clear impact on the mixing behavior of RMI-driven turbulence. 

Finally, we note that this work may be of particular interest to the development of modal models designed to initialize RANS turbulence models \cite{Rollin2013}. Such models must specify some set of modes to track, however a detailed assessment of best practices for specifying this initial set of modes remains to be done. Our work demonstrates the importance of the number and density of initial modes in 3D and also introduces the shocked heavy band problem.   We envision the shocked heavy band problem could be a relevant test case  to assess initialization procedures and benchmark predictions with current reduced-dimension (1D, 2D) RANS of turbulent mixing.

In conclusion, we have demonstrated that for shock-driven turbulent mixing the choice of spectral density of the initial perturbations has a clear effect on the mixedness achieved -- particularly at late times after reshock. Our analysis indicates that this effect is largely the result of an inhibited turbulent cascade for cases with low initial spectral density. In contrast, the total kinetic energy and mix width were not significantly impacted by changing the initial spectral density.

% Surround figure environment with turnpage environment for landscape
% figure
% \begin{figure}
% \end{figure}

% tables should appear as floats within the text
% The ruledtabular enviroment adds doubled rules to table and sets a
% reasonable default table settings.
% Use the table* environment to get a full-width table in two-column
% Add \usepackage{longtable} and the longtable (or longtable*}
% environment for nicely formatted long tables. Or use the the [H]
% placement option to break a long table (with less control than 
% \begin{table}%[H] add [H] placement to break table across pages
% \begin{ruledtabular}
% Lines of table here ending with \\
% \end{tabular}
% \end{ruledtabular}
% \end{table}

% Surround table environment with turnpage environment for landscape
% table
% \begin{table}
% \begin{ruledtabular}
% \end{tabular}
% \end{ruledtabular}
% \end{table}

%\appendix
%\section{}

\begin{acknowledgments}
We thank Jon Reisner and Betrand Rollin for their helpful contributions to the formulation of this work, and Ray Ristorcelli for suggesting improvements in our analysis. NJN is supported by a Nicholas C.~Metropolis Post-Doctoral Fellowship as part of the Advanced Simulation Capabilities Initiative at Los Alamos National Laboratory. Los Alamos National Laboratory is operated by Los Alamos National Security, LLC for the U.S.~Department of Energy, NNSA under contract No.~DE-AC52-06NA25396.
\end{acknowledgments}

%\clearpage
\bibliography{ShockMixing}

\end{document}